\documentclass[prd,aps,nofootinbib,showpacs,10pt]{revtex4}
\usepackage{amsmath,graphicx,epsfig,amssymb,dsfont,mathtools}

\usepackage[usenames]{color}
\usepackage{ulem} 
\usepackage{bigstrut}

\newcommand{\be}{\begin{eqnarray}}
\newcommand{\ee}{\end{eqnarray}}
\newcommand{\non}{\nonumber \\}

\begin{document} 
\title{Chiral Dynamics  and 
S-wave Contributions in  Semileptonic B decays}
\author{Michael D\"oring $^a$, Ulf-G.  Mei\ss ner $^{a,b}$ and   Wei Wang $^a$
}

\affiliation{ 
  $^a$ Helmholtz-Institut f\"ur Strahlen- und Kernphysik and Bethe Center for
Theoretical Physics, Universit\"at Bonn, D-53115 Bonn, Germany\\
$^b$ Institute for Advanced Simulation, Institut f\"ur Kernphysik and J\"ulich
Center for Hadron Physics, JARA-FAME and JARA-HPC, Forschungszentrum J\"ulich,
D-52425 J\"ulich, Germany}

\begin{abstract}
The flavor-changing neutral current process  $b\to s l^+l^-$ is  beneficial to
testing the standard model and hunting for new physics scenarios. In exclusive
decay modes like $B\to K^*(892)l^+l^-$,  the S-wave effects may not be
negligible and thus have to be  reliably estimated.  Using the scalar form
factors  derived from dispersion relations in two channels and matched to Chiral
Perturbation Theory,  we investigate the S-wave  contributions in  $\overline
B^0\to  K^- \pi^+ l^+l^-$, with the $K\pi$  invariant mass lying in the vicinity
of the mass of $K^*(892)$, and the $B_s\to    K^- K^+ l^+l^-$ with $m_{KK}\sim
m_{\phi}$.   We find that  the S-wave will modify differential decay widths  by 
about  $10\%$ in the process of $\overline B^0\to  K^- \pi^+ l^+l^-$ and about 5\% in
$B_s\to   K^- K^+ l^+l^-$.  A forward-backward asymmetry  for the charged kaon
in the final state arises from the interference between the S-wave and P-wave
contributions.  The measurement of this asymmetry offers  a  new way to
determine the variation of the  $K\pi$ S-wave phase versus the invariant  mass.  
\end{abstract}
\pacs{13.20.He; 14.40.Be;}
\maketitle


\section{Introduction}
\label{section:introduction}


One of the foremost open questions in our current knowledge of particle physics
is whether   new degrees of freedom are relevant for the phenomena at the TeV
energy scale. The quest for these new particles which have distinguishable 
signatures compared to the standard model (SM) particles are of high priority at
the ongoing collider experiments. On the other hand,  low-energy processes
may be influenced by them and the corresponding predictions on physical
observables will be shifted away from the SM. Rare $B$ decays, with highly
suppressed  decay probabilities in the SM, are likely sensitive to these new
degrees of freedom and therefore  can be exploited as indirect searches.  In
particular,  the flavor-changing-neutral-current processes  $B\to
K^*(892)l^+l^-$ and $B_s\to \phi(1020)l^+l^-$ can provide a wealth of
information, in terms of a number of observables ranging from  decay fractions,
forward-backward asymmetries (FBAs), polarizations to a full angular
analysis~\cite{Lees:2012tva,Wei:2009zv,Aaltonen:2011cn}.    The recent
measurements of  $B\to K^*(892)l^+l^-$  by the LHCb collaboration  based on the 
$1 fb^{-1}$~\cite{Aaij:2013iag} data sample  show no significant deviations from
the  SM
theory~\cite{Ali:1999mm,Chen:2002bq,Kruger:2005ep,Egede:2008uy,Altmannshofer:2008dz,Chiang:2009dx,Khodjamirian:2010vf,Bobeth:2011gi,Bobeth:2012vn,Jager:2012uw,Descotes-Genon:2013vna}.
This great success of the SM implies that  new physics effects are likely 
small, and therefore renders the precision predictions for   the involved
quantities particularly important.

The process $B\to K^*(892)(\to K\pi)l^+l^-$ is a quasi-four-body decay, and  in
principle   other  $K\pi$ resonant and nonresonant states may also  contribute
in the same final state, and thus dilute the discrimination between new physics
and standard model.  Theoretically,  a general formula that  includes various
contributions has  been derived in Refs.~\cite{Lu:2011jm,Li:2010ra}.  In terms
of helicity amplitudes,  a compact form for the full angular distributions is
obtained, through which the branching ratios, forward-backward asymmetries and
polarizations can be easily projected.  Adopting these formulas, the S-wave
contribution in the $B\to K\pi l^+l^-$ has been estimated in recent
publications~\cite{Becirevic:2012dp,Matias:2012qz,Blake:2012mb}, and it is pointed out that  
contributions from  the S-wave $K\pi$ state are not negligible and should be
taken into account in future measurements.   It is noticeable that in these
studies, the S-wave $K\pi$ interaction is parametrized   in terms of  a
Breit-Wigner formula, which is  not justified especially for the broad scalar
meson $\kappa~\equiv~K^*_0(800)$. 
 
On the other hand,  the scattering of light mesons are basic processes in QCD
that deserve accurate measurements.   In the $K\pi$ system, $S$-wave
interactions proceeding through  isospin $I=1/2$ states are of particular
interest because  they  depend on the presence of scalar resonances. Studies of
the  scalar meson $\kappa$  can thus benefit from   accurate measurements of the
$I=1/2$ $S$-wave phase below  $m_{K\pi}=1$ GeV.  The current available 
information on the $I=1/2$ $K\pi$ scattering comes from the  {}{decays}  $D^+\to
K^-\pi^+ e^+\nu_e$~\cite{delAmoSanchez:2010fd} or $D\to K\pi
\pi$~\cite{Link:2009ng,Aitala:2005yh,Oller:2004xm,Bediaga:2004bc,Magalhaes:2011sh,Kamano:2011ih} as well as $J/\psi$
decays~\cite{Bugg:2005ni}. For energies beyond $\sqrt{s}\sim 1$~GeV there is the
partial wave amplitude based on LASS data widely used in the
literature~\cite{Aston:1987ir}. First data from lattice calculation start to
emerge~\cite{Lang:2012sv,Beane:2006gj} though still at unphysical pion masses;
constraints from chiral perturbation theory can be used to extract the $K\pi$
phase from the finite volume used in lattice
calculations~\cite{Doring:2011nd,Doring:2012eu,Doring:2011vk}.  In the case of
$K\bar K$, the constraint from  $D_s\to K^+K^- e^+\nu_e$ is less
precise~\cite{Aubert:2008rs}, while the semileptonic  $D_s$ decays have been
used to examine the $\pi\pi$ interaction~\cite{Ecklund:2009aa}. 

In this work, we improve the analysis in Refs.~\cite{Becirevic:2012dp,Matias:2012qz,Blake:2012mb} by combining the
knowledge of B meson weak decays, mainly based on operator product expansion
and  perturbation theory in QCD, and the low-energy effective theory for
$K\pi/K\bar K$ interaction, namely  Chiral Perturbation Theory (CHPT).   CHPT
can be used to constrain the form factor. Multiple subtractions for the
Muskhelishvili-Omn\`es problem allow the systematic matching to CHPT and thus to
fix the polynomial ambiguity in $s$ to a given order. We pursue this method to
determine the $K\pi$ form factor by a match up to next-to-leading order. Also,
we present a numerical scheme to solve the set of integral equations for cases
when iteration does not converge. For the $K\bar K$ form factor, we rely on the
model predictions of unitarized CHPT at leading order. The methodology we
employ here was pioneered for B-decays in Ref.~\cite{Gardner:2001gc}.

Using these results for the  $K\pi/K\bar K$  scalar form factors and  the
heavy-to-light transition matrix elements calculated in   the perturbative QCD
approach~\cite{Keum:2000ph,Keum:2000wi,Lu:2000em,Lu:2000hj,Li:2012nk}, we study
the S-wave contribution  and its interference  with P-wave. We will show that
the size of the S-wave pollution to the differential decay width in $B\to
K^*l^+l^-$ is about  $10\%$ while it  is about 5\%  in $B_s\to \phi l^+l^-$. 
At last, we will discuss a  subtraction method in the integration over
$m_{K\pi}^2$ and  $m_{K\bar K}^2$, which projects out the P-wave contributions
and suppresses the effects of the S-wave to less than 1\%.

The  paper is organized as follows.   
Sec.~\ref{sec:differentialdecaydistribution} recalls  the differential decay
distributions and the partially   integrated quantities in $B\to K^*l^+l^-$ and 
$B\to K^*_0l^+l^-$. In Sec.~\ref{sec:lineshape}, we calculate   the necessary
$K^-\pi^+$ and $K^+K^-$ scalar form factors.  Sec.~\ref{sec:results}  contains 
our numerical predictions. We conclude in the last section. The form factors
calculated in the perturbative QCD approach,  the helicity decay amplitudes and normalization 
are relegated to the Appendices A,  B and C respectively. 


\section{$B\to K\pi l^+l^-$  angular distributions   } 
\label{sec:differentialdecaydistribution}

\begin{figure}\begin{center}
\includegraphics[scale=0.4]{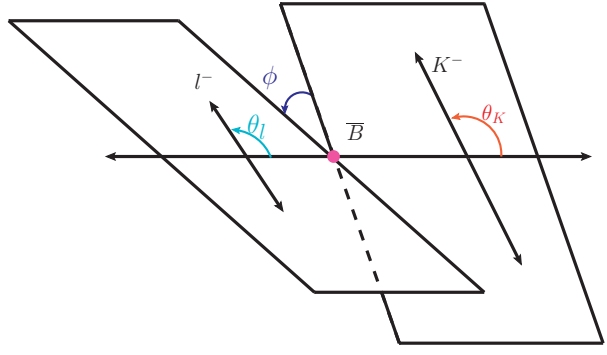}
\caption{Kinematics in  $\overline B\to  \overline K_J^*(\to K^-\pi^+)l^+l^-$. 
{The} $K_J^*$ moves along the $z$ axis in the $B$ rest frame. 
$\theta_K(\theta_l)$  is  the angle
between $z$-axis and the flight direction of $K^-$ ($\mu^-$)  in $K^*_J$ (lepton pair) rest frame, respectively.  
$\phi$ is the  azimuth angle  between the $K^*_J$ decay and lepton pair
planes. } 
\label{fig:angles}
\end{center}
\end{figure}

In this section we will  give  the angular distributions  for $B \to
K^-\pi^+l^+l^-$ (and   $B_s\to K^+K^-l^+l^-$ as well) with both S-wave and
P-wave contributions. Throughout this work, we will consider the $\bar B$ 
meson  and the pseudoscalar mesons in the final states to be charged [Neutral
mesons can be treated analogously].  We adopt  the  convention on the kinematics
in $B\to K_J^*(\to K\pi)l^+l^-$ as illustrated in Fig.~\ref{fig:angles}, where
$K^*_J$ is a generic kaon resonance with spin $J$. In the $B$ rest frame, the
$K_J^*$ flight direction  is chosen as the $z$ axis. $\theta_K$
($\theta_l$) is the polar angle  between the $K^-$ ($\mu^-$) moving
direction and the $z$ axis in the $K_J^*$ (lepton pair) rest frame. $\phi$ is
the angle between the two decay planes.

The effective Hamiltonian for the transition  $b\to sl^+l^-$  
 \begin{eqnarray}
 {\cal
 H}_{\rm{eff}}=
 -\frac{G_F}{\sqrt{2}}V_{tb}V^*_{ts}\sum_{i=1}^{10}C_i(\mu)O_i(\mu)
 \nonumber\label{eq:Hamiltonian}
 \end{eqnarray}
involves the  four-quark and the magnetic penguin operators $O_i$, and the $C_i(\mu)$
are the corresponding  Wilson coefficients for these local operators $O_i$. 
$G_F$ is the Fermi constant, $V_{tb}=0.999176$ and
$V_{ts}=-0.03972$~\cite{Beringer:1900zz} are the CKM matrix elements.  
The $b$ and $s$ quark masses are $m_b=(4.67^{+0.18}_{-0.06})$GeV and
$m_s=(0.101^{+0.029}_{-0.021})$GeV~\cite{Beringer:1900zz}.  The above effective
Hamiltonian results in the decay amplitude  
\begin{eqnarray}
 i {\cal M}(b\to
 sl^+l^-)&=&\frac{iG_F}{\sqrt2}\frac{\alpha_{\rm em}}{\pi}V_{tb}V_{ts}^*\times
 \left( \frac{C_9+C_{10}}{4}[\bar sb]_{V-A}[\bar ll]_{V+A}
 +\frac{C_9-C_{10}}{4}[\bar sb]_{V-A}[\bar ll]_{V-A}\right. \nonumber\\
 &&\left.+ C_{7L}m_b[\bar s i\sigma_{\mu\nu}
 (1+\gamma_5)b]\frac{q^\mu}{q^2}\times[\bar l \gamma^\nu l]
 + C_{7R}m_b[\bar s i\sigma_{\mu\nu}
 (1-\gamma_5)b]\frac{q^\mu}{q^2}\times[\bar l \gamma^\nu
 l]\right),\label{eq:decay-amplitude-bsll-LR}
\end{eqnarray}
where $C_{7L}=C_7$ and $C_{7R}=C_{7L}{m_s}/{m_b}$.

The process $B\to K_J^*(\to K\pi)l^+l^-$ is a four-body decay mode which undergoes
two steps in the resonance picture:  the $B$ meson first decays into a  excited kaonic state $K_J^*$ plus a pair of leptons; the $K_J^*$   propagates
followed by the strong decay into the $K\pi$.  Decay amplitudes of
$B\to (K\pi)l^+l^-$ can be obtained by sandwiching
Eq.~\eqref{eq:decay-amplitude-bsll-LR} between the initial and final hadronic states, in
which the spinor product $[\bar sb]$  will be replaced by hadronic matrix
elements defined in  Appendix A.  The operator realization of this picture may be
given as  
\begin{eqnarray}
 \langle l^+l^-| [\bar l l]|0\rangle  \langle K\pi | [\bar s b] |
 \overline B^0\rangle &\simeq & \langle l^+l^-| [\bar l l]|0\rangle  
 \int d^4 p_{K^*_J} \frac{ \langle K\pi|K^*_J\rangle 
 \langle K^*_J | [\bar s b] |\overline B^0\rangle }{ 
 p_{K^*_J}^2- m_{K^*_J}^2 +i m_{K^*_J} \Gamma_{K^*_J}},  
\end{eqnarray}
with $p^2_{K^*_J}=m^2_{K\pi}$.   In
Appendix~B, we collect  the required quantities  in terms  of  helicity
amplitudes that can lead to the full angular distributions
\begin{eqnarray}
 \frac{d^5\Gamma}{dm_{K\pi}^2dq^2d\cos\theta_K d\cos\theta_l d\phi}
 &=& \frac{3}{8}\Big[I_1^c  + 2I_1^s +(I_2^c +2I_2^s ) 
 \cos(2\theta_l) + 2I_3 \sin^2\theta_l
 \cos(2\phi)+2\sqrt 2I_4\sin(2\theta_l)\cos\phi \nonumber\\
 && +2\sqrt 2I_5 \sin(\theta_l) \cos\phi+2I_6 \cos\theta_l+2\sqrt 2I_7 
 \sin(\theta_l) \sin\phi\nonumber\\
 && +
 2\sqrt 2I_8\sin(2\theta_l)\sin\phi+2I_9 \sin^2\theta_l
 \sin(2\phi)\Big].
\end{eqnarray} 
In the massless limit for the involved leptons, and  integrating over the angles
$\theta_l,\theta_K$ and $\phi$, we have the dilepton mass spectrum
\begin{eqnarray}
\frac{ d^2\Gamma}{dq^2 dm_{K\pi}^2} 
 &\simeq &    |A^0_{L0}|^2+|A^0_{R0}|^2+|A^1_{L0}|^2+|A^1_{R0}|^2
 +  |A^1_{L\perp}|^2+|A^1_{L||}|^2+|A^1_{R\perp}|^2+|A^1_{R||}|^2 . 
\end{eqnarray}
where the functions $A_{L/Ri}$  are given by 
\begin{eqnarray}
 A_{L/R 0/t }&=& \sum_{J=0,1,2...}  A^J_{L/R 0/t }Y_{J}^0(\theta,0),\nonumber\\
 A_{L/R ||/\perp }&=& \sum_{J=0,1,2...}  A^J_{L/R ||/
 \perp }Y_{J}^{-1}(\theta,0),\nonumber \\
 A^J_{L/R 0/t }&=&   \sqrt{ N_{K_J^*}} {\cal M}_B(K^*_J, L/R, 0/t ) 
 L_{K^*_J}(m_{K\pi}^2) \equiv | A^J_{L/R 0/t }| 
 e^{i\delta^J_{{L/R 0/t }}},\nonumber\\
 A^J_{L/R ||/\perp }&=&   \sqrt{ N_{K_J^*}} {\cal M}_B(K^*_J, L/R, ||/\perp)
 L_{K^*_J}(m_{K\pi}^2)\equiv | A^J_{L/R ||/\perp}| e^{i\delta^J_{{L/R ||/
 \perp }}},\nonumber
\end{eqnarray}
with $ N_{K_J^*} = \sqrt{8/{3}} {\sqrt {\lambda}
{q^2}\beta_l}/({256\pi^3 m_B^3})$,  $\lambda\equiv
(m^2_{B}-m^2_{K_J^*}-q^2)^2-4m^2_{K_J^*}q^2$ and $\beta_l=
\sqrt{1-4m_l^2/q^2}$. The $K\pi$ lineshape is $L_{K^*_J}(m_{K\pi})$, and  for
the P-wave $K^*(892)$ we use the Breit-Wigner distribution:
\begin{eqnarray}
 L_{K^*_J}(m_{K\pi}^2)= \sqrt{ \frac{ m_{K^*_J} \Gamma_{K^*_J}  }{\pi} }
 \frac{1}{ p_{K^*_J}^2 -m_{K^*_J}^2+ i m_{K^*_J}\Gamma_{K^*_J}}.
\end{eqnarray}  
The handedness  label $L$ and $R$ denotes  the chirality of the di-lepton
system.  Expressions for the weak decay  amplitudes  $ {\cal M}_B$ can be found in 
Ref.~\cite{Lu:2011jm}, and to a good approximation, these amplitudes do not have
large strong phases. 

As a particular example,  we study the angular
distribution over $\theta_K$:  
\begin{eqnarray}
 \frac{d^3\Gamma}{ dq^2 dm_{K\pi}^2 d\cos\theta_K}    &\simeq&    
 \bigg\{    \frac{1}{2}  [|A^0_{L0}|^2+|A^0_{R0}|^2]  +   \sqrt {3} 
 \cos\theta_K  \big[  \cos(\delta_{L0}^0
 - \delta^{1}_{L0})|A^0_{L0}||A^{1}_{L0}| + \cos(\delta_{R0}^0 
 - \delta^{1}_{R0})|A^0_{R0}||A^{1}_{R0}|    \big] \nonumber\\
 &&    +  \frac{3}{2} \cos^2\theta_K     (|A^1_{L0}|^2+|A^1_{R0}|^2)       
 +  \frac{3}{4}\sin^2\theta_K  
 [     |A^1_{L\perp}|^2+|A^1_{L||}|^2+|A^1_{R\perp}|^2+|A^1_{R||}|^2    ]  
 \bigg\}.
 \label{eq:theta_K_distribution}
\end{eqnarray}
Compared to the distribution with only $B\to K^*(892)l^+l^-$,  the two terms in
the first line of Eq.~\eqref{eq:theta_K_distribution} are new: the first one is
the  S-wave $K\pi$ contribution,  while the second term  corresponds to  the
interference of   S-wave and P-wave.  Based on this interference,  one can  
define a forward-backward  asymmetry for the charged kaon, 
\begin{eqnarray}
 \frac{d^2A_{FB}^{K}}{ dq^2 dm_{K\pi}^2 } &\equiv &   \bigg[\int_{0}^1 
 - \int_{-1}^0\bigg] d\cos\theta_K \frac{d^3\Gamma}{ dq^2 dm_{K\pi}^2
  d\cos\theta_K}  \nonumber\\
 &=&   \sqrt {3} \big[  \cos(\delta_{L0}^0- 
 \delta^{1}_{L0})|A^0_{L0}||A^{1}_{L0}| +\cos(\delta_{R0}^0 - 
 \delta^{1}_{R0})|A^0_{R0}||A^{1}_{R0}|  \big].
\end{eqnarray}
 
The narrow-width  approximation is not valid as the S-wave $K\pi$  interaction
is strong. However, the Watson theorem implies that, in the elastic regime,
phases measured in $K\pi$ elastic scattering and in a decay channel in which the
$K\pi$ system has no strong interaction with other hadrons  are equal modulo
$\pi$ radians. When the
experimental data  {are} available, this ambiguity is solved by determining the
sign of the $S$-wave amplitude from data.  
At leading order in $\alpha_s$   the lepton pair $l^+l^-$ indeed decouple{s from} the 
$K\pi$ final state, and thus   phases of B to scalar $K\pi$  decay amplitudes  are equal to
$I=1/2$  $K\pi$ scattering with the same isospin and angular momentum.  As a consequence, we have
\begin{eqnarray}
 \langle (K\pi)_0 |\bar s \Gamma b|\bar B\rangle   
 \propto F_{K^-\pi^+}(m_{K\pi}),
\end{eqnarray}
with  $F_{K^-\pi^+}$ the scalar form factor. In this work, we will
approximately  use the perturbative QCD approach to compute  {the} $B\to K^*_0$
form factor, and  the line-shape is given as
\begin{eqnarray}
 L_{K^*_0}^{\rm \chi PT} (m_{K\pi}) 
 = {\cal N}_{\chi PT} F_{K^-\pi^+}(m_{K\pi}), 
 \label{eq:chiPTlineshape}
\end{eqnarray}
with ${\cal N}_{\chi PT}$ being the normalization constant  {evaluated in
Appendix~\ref{sec:normalization}}.   We will comment on  the Watson theorem and 
the use of perturbative QCD later. 

\section{S-wave Scalar Form Factors} 
\label{sec:lineshape}

Scalar $\pi\pi/K\bar K$ and $K\pi/K\eta$ form factors have been calculated
within a variety of approaches using (unitarized) chiral perturbation
theory~\cite{Gasser:1990bv,Meissner:2000bc,Oller:2000ug,Frink:2002ht,Bijnens:2003uy,Lahde:2006wr,Bernard:2009ds,Guo:2012yt}
and dispersion
relations~\cite{Donoghue:1990xh,Jamin:2000wn,Jamin:2001zq,Jamin:2006tj,Bernard:2007tk,Bernard:2009ds},
in many cases using the former to constrain polynomial ambiguities of the
latter. Data exist for the pion vector form
factor~\cite{Guerrero:1997ku,Guo:2008nc,Hanhart:2012wi} and the $\pi\pi$ scalar
form factor (e.g., the $\phi$ in $J/\psi\to \pi\pi\phi$ acts as an isospin
filter). However, the strangeness changing scalar form factor is more difficult
to extract due to the mixing of $S$ and $P$ waves.

In Ref.~\cite{Jamin:2001zq}, once-subtracted dispersion relations in the three
channels $K\pi$, $K\eta$, and $K\eta'$ were solved to determine the form factor.
Here, we solve the two-channel ($K\pi$, $K\eta$) problem, but with two
subtractions to match not only the value but also the slope to the corresponding
next-to-leading order expressions of CHPT. As it turns out, this allows for a
good prediction of the known value of the form factor at the Callan-Treiman
point. With an additional subtraction it would be possible to explicitly include
this point as a constraint in the relations. For the one-channel case, the
corresponding Omn\`es representation of the form factor has been formulated in
Ref.~\cite{Bernard:2011ae}. 

Though inelasticities from $K\eta$ are usually taken as small in the $\kappa$
channel, the $K\eta$ channel is present. To estimate its influence, we perform a
global fit to various $\pi\pi/K\bar K$ and $K\pi/K\eta$ scattering channels
using the Inverse Amplitude Method (IAM) in the formulation of
Ref.~\cite{Oller:1998hw}. The original fit result of Ref.~\cite{Oller:1998hw}
produced an unsatisfactory  description of the data in the $\kappa$ channel. The fit of
the low-energy constants was improved in Ref.~\cite{Doring:2011nd}. Here, we
extend the range of data description to higher energies, to take account of the
$K_0^*(1430)$, by means of a bare explicit $s$-channel resonance propagator,
dressed through the couplings to the $K\pi$ and $K\eta$ channels.
Coupled-channel unitarity is preserved. 

The resulting amplitude is then regarded as a representation of the phase-shift
data (including inelasticities) that serves as input for the 
Muskhelishvili-Omn\`es problem. For the solution of the latter we propose a
numerical method to directly invert the integral equations instead of solving
them by iteration. To check the influence of the inelasticity, in a second step
we perform a one-channel refit to the phase-shift data, dropping the constraints
from data other than the $\kappa$ channel. The one-channel problem is solved
both by direct inversion and with the Omn\`es function. 

For the $\pi\pi/K\bar K$ form factor evaluated in Sec.~\ref{sec:pipiff} we rely
on the prediction of the chiral unitary approach at order $p^2$, matched to the
NLO expression of the form factor at order $p^2$. It should be noted that a
corresponding procedure using the IAM in the formulation of
Ref.~\cite{Oller:1998hw}, that contains also the ${\cal O}(p^4)$ contact terms,
is not possible: in that formulation the method produces, almost unavoidably,
spurious singularities between $s=0$ and the lowest threshold for meson-meson
scattering. These singularities are then also present in the form factor.	


\subsection{Scalar $K\pi$ and $K\eta$ form factors}
\subsubsection{The $K\pi$ scattering amplitude}
\label{sec:scattering}

To obtain the coupled-channel form factor via the Muskhelishvili-Omn\`es
relations, the scattering amplitude $T$ needs to be known, parameterized in the
present study by the IAM plus a genuine resonance term,
\begin{eqnarray}
 T&=&(1-VG)^{-1}V,\quad V=V_{\rm IAM}+V_{\rm res},\non
 V_{\rm IAM}&=&\left(1-V^{[4]}(V^{[2]})^{-1}\right)^{-1}V^{[2]},\non
 (V_{\rm res})_{ij}
 &=&\frac{g_i\,g_j\left[s-(M_\eta+M_K)^2\right]^2}{f_\pi^2\left(s-m_b\right)^2}
\label{model}
\end{eqnarray}
with the matrices in channel space $V^{[2]}$ and $V^{[4]}$ containing the ${\cal
O}(p^2)$ and ${\cal O}(p^4)$ contact interactions of Ref.~\cite{Oller:1998hw}.
An explicit expression for the scalar loop function $G$ can be found, e.g., in
the Erratum of Ref.~\cite{Oller:1998hw}. The $T$-matrix is projected to
different partial waves, but the explicit resonance is only inserted in the
$\kappa$ channel to take account of the $K_0^*(1430)$; $t$ and $u$-channel
resonance contributions are neglected. The mixing of $\eta$-$\eta'$ is neglected
throughout this study; for example, in the recent study of meson-meson
scattering of Ref.~\cite{Guo:2011pa} it is taken into account. Another resonance
fit to $K\pi$ has been performed in Ref.~\cite{Jamin:2000wn}, see also
Refs.~\cite{Albaladejo:2008qa,GomezNicola:2001as}, and for an early study of
combining CHPT with resonances, see~\cite{Bernard:1991zc}.
  Fit parameters of the
present study are the seven low energy constants $L_1^r$ to $L_5^r$ and $L_7^r$,
$L_8^r$~\cite{Oller:1998hw} appearing in $V^{[4]}$, as well as the two bare
resonance couplings $g_1,\,g_2$ to the $K\pi$ and $K\eta$ channels,
respectively, and the bare mass $m_b$. 

\begin{figure}
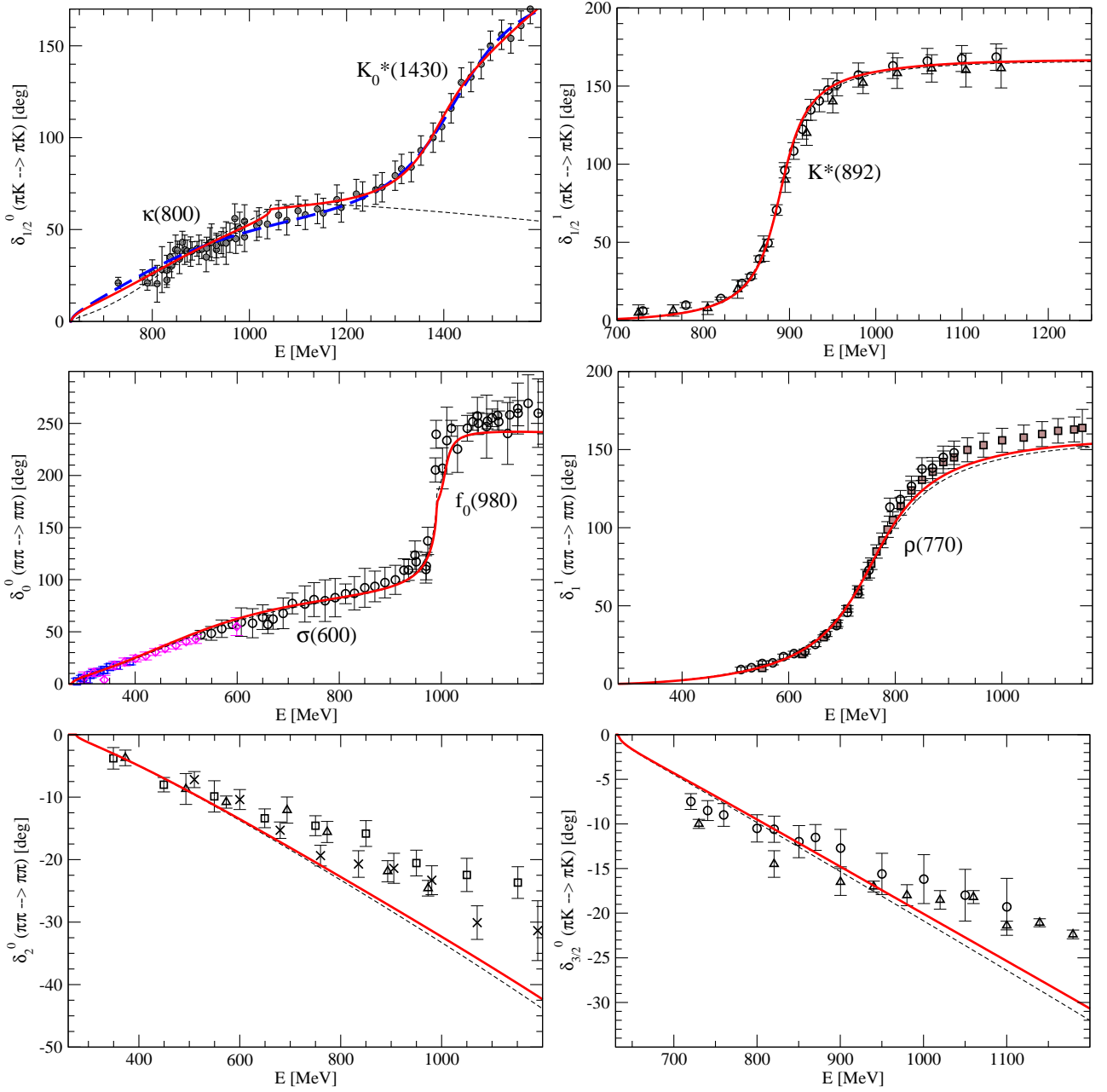
\begin{center}
\includegraphics[scale=0.45]{phase_shift_ILS_12_0_-1}
\includegraphics[scale=0.45]{phase_shift_ILS_12_1_-1}
\includegraphics[scale=0.45]{phase_shift_ILS_0_0_0}
\includegraphics[scale=0.45]{phase_shift_ILS_1_1_0}
\includegraphics[scale=0.45]{phase_shift_ILS_2_0_0}
\includegraphics[scale=0.45]{phase_shift_ILS_32_0_-1}
\caption{Solid (red) lines: Combined coupled-channel fit to $\pi\pi$ and $K\pi$
scattering in $S$- and $P$-waves. Thin (black) dashed lines: Results from
Ref.~\cite{Doring:2011nd}. Dashed (blue) line for the $\kappa$ channel:
One-channel re-fit (only to the data in the $\kappa$ channel).  Data: see
Ref.~\cite{Doring:2011nd} and references therein. The data for the $\kappa$
channel are from Ref.~\cite{Oller:1998hw} containing the data of
Refs.~\cite{Aston:1987ir} for the higher energies.} 
\label{fig:phase}
\end{center}
\end{figure}

The fit result for the considered partial waves and reaction channels is shown
in Figs.~\ref{fig:phase} with the (red) solid lines and compared to the previous
solution of Ref.~\cite{Doring:2011nd} (thin dashed lines).
The three new resonance parameters allow for a good data description in the
energy region of the $K_0^*(1430)$. Obviously, these new parameters introduce
additional freedom that is reflected in a slightly improved data description
also for the other considered channels and reactions. 

It should be noted that the amplitude for the $\kappa$ quantum numbers exhibits
a visible cusp from the $K\eta$ channel at $E=1.043$~GeV which is an indication
of considerable inelasticity. In the previous solution from
Ref.~\cite{Doring:2011nd} (thin dashed line), the cusp is even more pronounced.
We accept this behavior as a consequence of the combined fit to different
reactions, but will compare with a one-channel description in the following. For
that, the model of Eq.~(\ref{model}) is reduced to the $K\pi$ channel and only
the phase-shift data of the $\kappa$ channel is fitted for simplicity. This
means no attempt is undertaken to determine the low-energy constants, but the
fit serves merely as a one-channel representation of the phase-shift data. The
result is shown with the thick dashed (blue) line in Fig.~\ref{fig:phase}.

The values of the low energy constants of the global two-channel fit are close
to the ones quoted in Ref.~\cite{Doring:2011nd}. The same applies to the pole
positions of the $K^*(892)$, $\rho (770)$, $\sigma(600)$, and $f_0(980)$
resonances. Here, we quote only the pole positions and residues of the
$\kappa(800)$ and the $K_0^*(1430)$ to open channels. As Table~\ref{tab:popores}
shows
\begin{table}
\begin{center}
\noindent\begin{tabular}{lllr@{}lr@{}lr@{}l}
\hline \hline
 &		&    	& \multicolumn{2}{c}{$z_0$~[MeV]\hspace*{0.4cm}}& \multicolumn{2}{c}{$a_{-1} (K\eta)$~[$M_\pi$]}& \multicolumn{2}{c}{$a_{-1} (K\pi)$~[$M_\pi$]}\bigstrut\\
\hline
$\kappa(800)$	& this work &(2-ch.)\hspace*{0.4cm}	&  $792$ & $-i\,279$	       		&	&				& $-29$ & $-i\,57$	       \bigstrut[t]\\
		& this work &(1-ch.)			&  $715$ & $-i\,283$	       		& 	&				& $-45$ & $-i\,62$	       \\
		& Ref.~\cite{Doring:2011nd} &($\chi$U)	&  $815$ & $-i\,226$	       		& 	&				& $-30$ & $-i\,57$	       \\
	      	& Ref.~\cite{DescotesGenon:2006uk} &(Roy-S.)&  $658$ & $-i\,279$   	       		& 	&				& $$	& $$	               \\
$K_0^*(1430)$	& this work &(2-ch.)    		&  $1388$& $-i\,71$	       		& $-11$ & $-i\,5$			& $11$  & $+i\,13$	       \\
		& this work &(1-ch.)			&  $1425$& $-i\,120$			& $0$ 	&  				& $20$  & $+i\,39$	       \\
	      	& Ref.~\cite{Bugg:2009uk}& (phen.)	&  $1427$& $-i\,135$   	       		& 	&				& $$	& $$		       \bigstrut[b]\\
\hline \hline
\end{tabular}
\end{center}
\caption{The $\kappa(800)$ and $K_0^*(1430)$ pole positions $z_0$~[MeV] and
residues $a_{-1}[M_\pi]$ to open channels. Uncertainties quoted in other works
have been suppressed.}
\label{tab:popores}
\end{table}
the results for both resonances depend indeed significantly on whether a one- or
a two-channel system is considered. In the two-channel (2-ch.) calculation the
$\kappa(800)$ pole is at higher energies than in the one-channel calculation
(1-ch.) in which case the result is closer to the Roy-Steiner determination
(Roy-S.) of Ref.~\cite{DescotesGenon:2006uk}, see also
Refs.~\cite{Buettiker:2003pp,DescotesGenon:2007ta}. For the $K_0^*(1430)$, the
one-channel calculation results in a broader resonance close to the one-channel
phenomenological analysis (phen.) of Ref.~\cite{Bugg:2009uk}, while the
$K_0^*(1430)$ is much narrower in the two-channel calculation (with around 35\%
branching fraction into $K\eta$).  In the latter case, the pole position is in
agreement with a recent three-channel chiral unitary analysis~\cite{Guo:2011pa}
in which a pole position of $z_0=1428^{+56}_{-23}-i\,87^{+53}_{-28}$~[MeV] is
quoted. 


\subsubsection{The $K\pi$ form factor}

The strangeness-changing scalar form factors are defined as
\begin{eqnarray}
 \langle 0| \bar su |K\pi\rangle = \frac{m_K^2- m_\pi^2} {m_s-m_u} C_X F_X(s). 
\label{defff}
\end{eqnarray}
with $C_X$ being the normalization constants  
\begin{eqnarray}
 C_{K^+\pi^0} =\frac{1}{\sqrt 2}, \;\; C_{K^0\pi^+}=1, \;\; 
 C_{K^+ \eta_8}= -\frac{1}{\sqrt 6},\;\; C_{K^+\eta_1}= \sqrt{\frac{4}{3}}. 
\end{eqnarray}
The $K\pi$ and $K\eta_8$ form factors $f_{K\pi}$ and $f_{K\eta_8}$ have been
calculated to NLO in Ref.~\cite{Gasser:1984ux}. To project to isospin $I=1/2$,
the corresponding coefficients have to be determined. For that, we proceed in
analogy to Ref.~\cite{Meissner:2000bc}. Here, the strangeness-changing
combinations can be obtained from a source term
\be
\chi_s=2B_0\left(
\begin{array}{ccc}
0		& 0		& m_{\bar us}\\
0		& 0		& m_{\bar ds}\\
m_{\bar su}	& m_{\bar sd}	& 0
\end{array}
\right)
\ee
with the basis $(u,d,s)$. In
the lowest-order chiral effective Lagrangian, the scalar source appears as
\be
{\cal L}=\frac{f^2}{4}\langle U^\dagger\chi_s+\chi_s^\dagger U\rangle
\label{chil}
\ee
and, e.g., $\displaystyle{\bar su=- {\partial {\cal L}}/{\partial m_{\bar
su}}}$. To two meson fields in Eq.~(\ref{chil}), the terms arising from all
strangeness changing terms $m_{\bar qq'}$ are (we quote only the sum):
\be
\bar qq'=\bar us+\bar ds+\bar su+\bar sd
&=&\frac{B_0}{6}
\Big[6\left(\pi^-K^++K^-\pi^++\pi^+K^0+\pi^-\bar K^0\right)\non
&-&\sqrt{2}\left(K^-+K^+\right)\left(\sqrt{3}\eta_8-3\pi^0\right)\non
&-&\sqrt{2}\left(K^0+\bar K^0\right)\left(\sqrt{3}\eta_8+3\pi^0\right)\Big] \ .
\ee
The isospin $I=1/2$ combination is 
\be
\langle 0|\bar qq'|K\pi\rangle_{I=1/2} =\sqrt{\frac{3}{2}}\,B_0\ , \quad
\langle 0|\bar qq'|K\eta_8\rangle_{I=1/2}&=&-\frac{1}{\sqrt{6}}\,B_0 \ .
\ee
In the following, we omit an overall factor of $\sqrt{3/2}B_0$ and obtain
the form factors in isospin $I=1/2$ as
\be
F_{K\pi }=f_{K\pi},\quad F_{K\eta}=-\frac{1}{3}\,f_{K\eta_8} \ .
\ee
The overall normalization of the form factors is estimated in
Appendix~\ref{sec:normalization}. Recasting the expressions for $f_{K\pi}$ and
$f_{K\eta_8}$ from Ref.~\cite{Gasser:1984ux} in terms of the leading-order
meson-meson scattering transitions $K$ we obtain
\begin{eqnarray}
F^\chi_{K\pi }(s)&=& 
1+\frac{4L_5^r\,s}{f^2}+\frac{s}{4\Delta_{K\pi}}
\left(5\mu_\pi-2\mu_K-3\mu_{\eta_8}\right)
+\bar J_{K\pi}K_{K\pi,K\pi}-\frac{1}{3}\,
\bar J_{K\eta_8}K_{K\eta_8,K\pi} \ , 
\non
F^\chi_{K\eta_8}(s)&=&-\frac{1}{3}-\frac{4L_5^r\,s}{3f^2}
-\frac{3s}{4\Delta_{K\pi}}\left(\mu_\pi-2\mu_K+\mu_{\eta_8}\right)
+\bar J_{K\pi}K_{K\pi,K\eta_8}-\frac{1}{3}\,
\bar J_{K\eta_8}K_{K\eta_8,K\eta_8} \ ,
\label{ffchi}
\end{eqnarray}
where 
\begin{eqnarray}
&&K_{K\pi ,K\pi} =-\frac{1}{8 f^2}\left(2\Sigma-5s
+\frac{3\Delta_{K\pi}^2}{s}\right),\quad 
K_{K\pi ,K\eta_8}=-\frac{1}{8 f^2}\left(3s-2\Sigma
-\frac{\Delta_{K\pi}^2}{s}\right) \ ,
\non 
&&K_{K\eta_8,K\eta_8}=-\frac{1}{24f^2}\left(9s+\frac{\Delta_{K\pi}^2}{s}
-18M_{\eta_8}^2-2M_K^2\right) \ ,
\end{eqnarray}
and $K_{K\eta_8,K\pi}=K_{K\pi ,K\eta_8}$ due to time reversal invariance,
$\Sigma=M_\pi^2+M_K^2$, $\Delta_{K\pi}=M_K^2-M_\pi^2$ and
\be
\bar J&=&\frac{1}{32\pi^2}\bigg[2+\left(\frac{M_1^2-M_2^2}{s}
-\frac{M_1^2+M_2^2}{M_1^2-M_2^2}\right)\log\frac{M_2^2}{M_1^2}
-\frac{\lambda(s)}{s}\bigg(\log(s+\lambda(s)+M_1^2-M_2^2)
+\log(s+\lambda(s)-M_1^2+M_2^2)\non
&&-\log(-s+\lambda(s)-M_1^2+M_2^2)-\log(-s+\lambda(s)+M_1^2-M_2^2)\bigg)\bigg].
\ee
Note the convention $V^{[2]}=-K$ with $V^{[2]}$ from Eq.~(\ref{model}). The
constant $f$ is taken to equal the pion decay constant  $f_\pi = 92.4$~MeV,
$\lambda^2(s)=[s-(M_1+M_2)^2][s-(M_1+M_2)^2]$, and $s\equiv s+i\epsilon$ ensures
that the correct sheet of the logarithm is taken. The expressions for  the
logarithms $\mu_i$ generated by chiral tadpoles in the NLO  scalar form factors
are given by
\begin{equation}
\mu_i = \frac{M_i^2}{32\pi^2 f^2} \log\left(\frac{M_i^2}{\mu^2}\right).
\label{tadp}
\end{equation}
For the low-energy constant $L_5^r$ we take the value from
Ref.~\cite{Bijnens:2011tb} of the fit {\it All $p^4$},
$10^3\cdot\,L_5^r(\mu=M_\rho)=1.21$. In the present calculation the scale is set
by the cutoff of the scalar loop function $G$ in Eq.~(\ref{model}) of
$\Lambda=1$~GeV.  The scale can be calculated according to
Ref.~\cite{Oller:1998hw} 
\be
\mu'=\frac{2\Lambda}{\sqrt{e}}
\label{matching}
\ee
up to order ${\cal O}(M_i^2/\Lambda^2)$. This expression also holds for loops
with unequal masses, as we have checked explicitly. The value of the low energy
constant at this $\mu'$ is obtained by evolving $L_5^r$ from $\mu=M_\rho$ using
the $\beta$-function~\cite{Gasser:1984gg},
\be
L_5^r(\mu)-L_5^r(\mu')=\frac{3}{8\,(4\pi)^2}\log\frac{\mu'}{\mu} \ ,
\ee
resulting in $10^3\cdot\,L_5(\mu'=1.213\,{\rm GeV})=0.131$. 


\subsubsection{Matching to CHPT by twice-subtracted dispersion relations}
\label{sec:solu}

We write the twice-subtracted Muskhelishvili-Omn\`es problem for the two channels $K\pi$ and $K\eta$ as
\begin{eqnarray}
F_{K\pi }(s)&=&F^\chi_{K\pi}(0)+(F^\chi_{K\pi})'(0)\,s
    +\frac{s^2}{\pi}\int\limits_{s_{K\pi }}^\infty ds'\,
    \frac{F_{K\pi }(s')\,\sigma_{K\pi }(s')\,
    T^*_{K\pi ,K\pi }(s')}{s'^2(s'-s-i\epsilon)}
    +\frac{s^2}{\pi}
    \int\limits_{s_{K\eta}}^\infty ds'\,\frac{F_{K\eta}(s')\,
    \sigma_{K\eta}(s')\,
    T^*_{K\eta,K\pi }(s')}{s'^2(s'-s-i\epsilon)} \ ,
    \non
F_{K\eta}(s)&=&F^\chi_{K\eta}(0)+(F^\chi_{K\eta})'(0)\,s
    +\frac{s^2}{\pi}\int\limits_{s_{K\pi }}^\infty ds'\,
    \frac{F_{K\pi }(s')\,\sigma_{K\pi }(s')\,T^*_{K\eta,K\pi }(s')}{
    s'^2(s'-s-i\epsilon)}
    +\frac{s^2}{\pi}\int\limits_{s_{K\eta}}^\infty ds'\,\frac{F_{K\eta}(s')\,
    \sigma_{K\eta}(s')\,T^*_{K\eta,K\eta}(s')}{s'^2(s'-s-i\epsilon)}  \non
\label{mo}
\end{eqnarray}
with $T$ from Eq.~(\ref{model}) and\footnote{The matrix $T$ is defined with
opposite sign in Ref.~\cite{Jamin:2001zq} which is here absorbed in the sign of
$\sigma$.} $\sigma=-q_{\rm c.m.}/(8\pi\sqrt{s})$. The subtractions ensure that
the scalar form factors $F_{K\pi }$ and $F_{K\eta}$ match the size and slope of
the next-to-leading order chiral result from Eq.~(\ref{ffchi}) at $s=0$.

The system of integral equations (\ref{mo}) cannot be solved by iteration as it
is possible in other cases~\cite{Jamin:2001zq,Niecknig:2012sj}. Due to the two
subtractions, this procedure is not convergent. However, it can be solved by
matrix inversion. The derivation of the solution is shown for the one-channel
case and then generalized to the two-channel case. We rewrite the twice
subtracted dispersion relation to numerically regularize the singularity
($s\equiv s+i\epsilon$),
\be
F(s)&=&F^\chi(0)+(F^\chi)'(0)\,s+K(s,s)F(s)\left[\log\left(
\frac{s_{{\rm cut}}-s}{s}\right)+i\pi\right]
    +\int\limits_{0}^{s_{{\rm cut}}} ds'\,
    \frac{K(s,s')F(s')-K(s,s)F(s)}{s'-s-i\epsilon}\non
\label{regu}
\ee
with $s_{{\rm cut}}\to \infty$ and the kernel
\be
 K(s,s')=\frac{s^2}{\pi}\frac{\sigma(s')T^*(s')}{s'^2}\, 
 \Theta(s'-s_{{\rm thres.}})\ .
\ee
Adding and subtracting the singularity in this way ensures that for $s>s_{{\rm
thres.}}$ the integral is regular and the imaginary part is correctly evaluated
while for $s<s_{{\rm thres.}}$ these extra terms are absent. Additionally, the
lower integration limit has been moved to the definition of the kernel $K$, the
reason for which will become clearer in the following.

For the numerical evaluation, the integral is replaced by a sum according to
$\int ds' f(s')\to \sum_j w_j f(s'_j)$ with integration weights $w_j$ at
$s'=s'_j$. To apply the matrix inversion technique, one has to choose
$s_j=s'_j\, , \forall j=1,\dots,n$, such that Eq.~(\ref{regu}) is rewritten in
discretized form as
\be
F(s_i)&=&F^\chi(0)+(F^\chi)'(0)\,s_i+K(s_i,s_i)\left[\log\left(
\frac{s_{{\rm cut}}-s_i}{s_i}\right)+i\pi\right]
\non
&+&w_i\,\frac{K(s_i,s_{i+1})F(s_{i+1})-K(s_i,s_i)F(s_i)}{s_{i+1}-s_i}
+\sum_{j\neq i}^n w_j\,\frac{K(s_i,s_j)F(s_j)-K(s_i,s_i)F(s_i)}{s_j-s_i}
\quad \forall i=1,\dots,n \ .
\label{finalnum}
\ee
The sum over the integration weights has been split into a regular part $i\neq
j$, and the term $i=j$ has been replaced by the right-hand derivative, 
\be
 \lim_{s'\to s}\left(\frac{K(s,s')F(s')-K(s,s)F(s)}{s'-s}\right)
 \to \frac{K(s_i,s_{i+1})F(s_{i+1})-K(s_i,s_i)F(s_i)}{s_{i+1}-s_i} \ .
\ee
For the case $i=n$, this term is simply set to zero which induces an error that
vanishes for sufficiently many integration points; furthermore, the derivative
should be small for large $s$. One could introduce here a left-hand derivative,
but this unnecessarily complicates the equations.
Introducing
\be
M_{ij}=
\begin{dcases*}
\frac{K(s_i,s_j)}{s_j-s_i}\,w_j 				& for $j\neq i$, $j\neq i+1$,\\
\frac{K(s_i,s_{i+1})}{s_{i+1}-s_i}\,\left(w_i+w_{i+1}\right) 	& for $j= i+1$, $i\neq n$, \\
K(s_i,s_i)\left[\log\left(\frac{s_{{\rm cut}}-s_i}{s_i}\right) +i\pi-\frac{w_i}{s_{i+1}-s_i}
-\sum_{j\neq i}\frac{w_j}{s_j-s_i}\right]			& for $i=j\neq n$, \\
K(s_i,s_i)\left[\log\left(\frac{s_{{\rm cut}}-s_i}{s_i}\right) +i\pi
-\sum_{j\neq i}\frac{w_j}{s_j-s_i}\right]			& for $i=j=n$ \ ,
\end{dcases*}
\label{mij}
\ee
and the vector
\be
 {\bf F}^\chi=\Big(F^\chi(0)+(F^\chi)'(0)\,s_1,\dots,
 F^\chi(0)+(F^\chi)'(0)\,s_n\Big)^T \ ,
 \label{inh}
\ee 
Eq.~(\ref{finalnum}) is rewritten as
\be
{\bf F}={\bf F}^\chi+{\bf M}\,{\bf F}
\ee
that can be inverted to yield the form factor 
\be
{\bf F}=(\mathds{1}-{\bf M})^{-1}\,{\bf F}^\chi \ .
\ee
The scheme presented here resembles the Haftl-Tabakin scheme used for the
solution of scattering problems~\cite{Haftel:1970zz}. However,
the present scheme is more involved, because one has explicit singularities for
$s=s'$. As shown, those can be handled by using the numerical derivative of the
integrand. In the context of a dispersive analysis of the scalar form factor of
the nucleon, a similar method has been derived in
Ref.~\cite{Hoferichter:2012wf}. 

The generalization to two or more channels is straightforward. For this, we
introduce the channel indices $\mu,\nu$ in the kernel  
\be
K_{\mu\nu}(s,s')=\frac{s^2}{\pi}\frac{\sigma_\nu(s')T^*_{\nu\mu}(s')}{s'^2}\, \Theta(s'-s_{{\rm thres. }\nu})
\label{kernel2}
\ee
that lead to a $2n\times 2n$ matrix ${\bf M}_{\mu\nu}$ according to
Eq.~(\ref{mij}). The initial coupled-channel problem of Eq.~(\ref{mo}) can
finally be rewritten as
\be
\left(
\begin{array}{l}
{\bf F}_1\\
{\bf F}_2
\end{array}
\right)
=
\left(
\begin{array}{l}
{\bf F}^\chi_1\\
{\bf F}^\chi_2
\end{array}
\right)
+
\left(
\begin{array}{ll}
{\bf M}_{11}&{\bf M}_{12}\\
{\bf M}_{21}&{\bf M}_{22}
\end{array}
\right)
\,
\left(
\begin{array}{l}
{\bf F}_1\\
{\bf F}_2
\end{array}
\right)
\label{fintwo}
\ee
where 1 stands for the $K\pi$ channel and 2 for the $K\eta$ channel. The
solution for the $2n$ elements
$(F_{K\pi}(s_1),\dots,F_{K\pi}(s_n),F_{K\eta}(s_1),\dots,F_{K\eta}(s_n))^T$ is
then obtained by matrix inversion as before~\footnote{It finally becomes also
clear why we have extended the lower integration limit to zero in writing
Eq.~(\ref{regu}). In this way, coupled-channel problems with different
thresholds can be treated easily, and, second, the solution obtained for the
form factors extends down to $s=0$.}.


\subsubsection{Results for the form factors}

\begin{figure}\begin{center}
\includegraphics[scale=0.5]{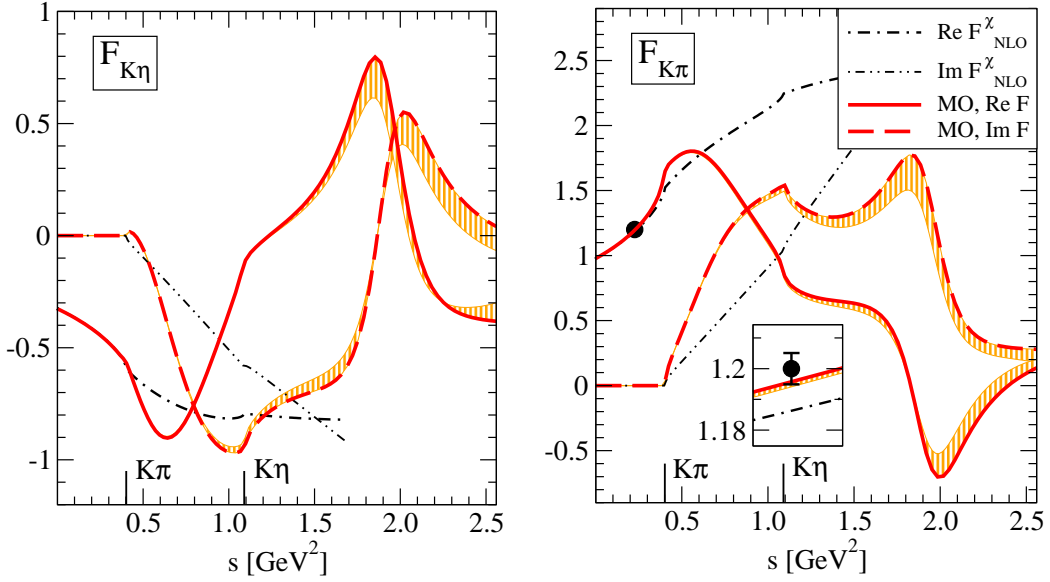}
\caption{The $K\eta$ (left panel) and $K\pi$ (right panel) form factors. Thick (red) lines:
Muskhelishvili-Omn\`es solution according to Eq.~(\ref{mo}), obtained with
Eq.~(\ref{fintwo}). Also, the uncertainties from higher energy input are shown
(hatched areas). Dash-dotted lines: chiral next-to-leading order
result~\cite{Gasser:1984ux} according to Eq.~(\ref{ffchi}). The data point shows
the value at the Callan-Treiman point (not included as a constraint).} 
\label{fig:result_FF}
\end{center}
\end{figure}

With the $T$ matrix of Eq.~(\ref{model}) and the chiral form factors of
Eq.~(\ref{ffchi}) as input, the solution of Eqs.~(\ref{mo}) is shown in
Fig.~\ref{fig:result_FF} for $s_{{\rm cut}}=(2.05\,\text{GeV})^2$. The error
bands are obtained by varying $s_{{\rm cut}}$ from this value down to $s_{{\rm
cut}}=(1.65\,\text{GeV})^2$ to show the sensitivity to the high-energy input. As
the figure shows, values and slopes of the chiral results $F_{K\pi}^\chi(0)$ and
$F_{K\eta}^\chi(0)$ are matched. We have checked that $F_{K\pi}$ fulfills
Watson's theorem between the $K\pi$ and the $K\eta$ threshold. An additional
piece of information is given by the value of the form factor at the
Callan-Treiman point~\cite{Jamin:2006tj,Bernard:2011ae},
$F_{K\pi}(\Delta_{K\pi})=1.2346(53)\,F_{K\pi}(0)$. It is possible to explicitly
include this value in the Omn\`es representation of the form
factor~\cite{Bernard:2011ae}, fixing not only its value and slope at $s=0$ but
also the curvature. In the inset of Fig.~\ref{fig:result_FF}, we simply show our
prediction of the value.

\begin{figure}\begin{center}
\includegraphics[scale=0.5]{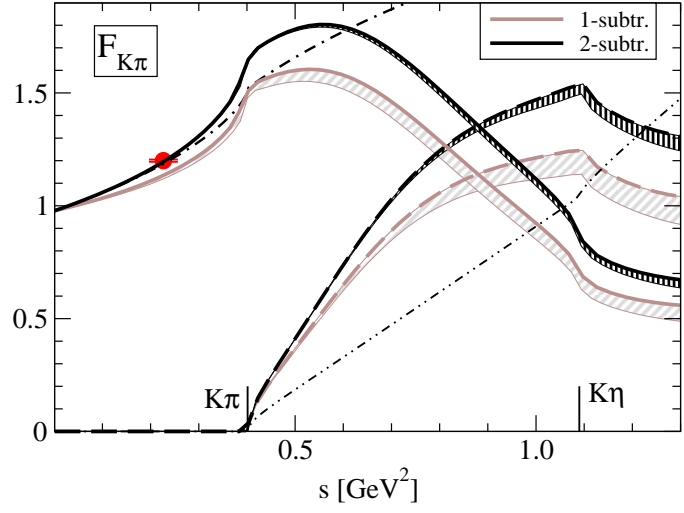}
\caption{Comparison of the once-subtracted Muskhelishvili-Omn\`es solution
(faint gray lines) with the twice-subtracted result of Fig.~\ref{fig:result_FF}
(black lines). Other curves as in Fig.~\ref{fig:result_FF}. For the
once-subtracted solution, the slope at $s=0$ is not matched and uncertainties
from higher-energy input (shaded areas) are larger.} 
\label{fig:comp_once}
\end{center}
\end{figure}
In Fig.~\ref{fig:comp_once} we compare to the once-subtracted version of the
Muskhelishvili-Omn\`es problem, given by
\begin{eqnarray}
F_{K\pi }(s)&=&F^\chi_{K\pi}(0)
    +\frac{s}{\pi}\int\limits_{s_{K\pi }}^\infty ds'\,
    \frac{F_{K\pi }(s')\,\sigma_{K\pi }(s')\,
    T^*_{K\pi ,K\pi }(s')}{s'(s'-s-i\epsilon)}
    +\frac{s}{\pi}\int\limits_{s_{K\eta}}^\infty ds'\,\frac{F_{K\eta}(s')\,
    \sigma_{K\eta}(s')\,T^*_{K\eta,K\pi }(s')}{s'(s'-s-i\epsilon)} \ , \non
F_{K\eta}(s)&=&F^\chi_{K\eta}(0)
    +\frac{s}{\pi}\int\limits_{s_{K\pi }}^\infty ds'\,
    \frac{F_{K\pi }(s')\,\sigma_{K\pi }(s')\,
    T^*_{K\eta,K\pi }(s')}{s'(s'-s-i\epsilon)}
    +\frac{s}{\pi}\int\limits_{s_{K\eta}}^\infty ds'\,
    \frac{F_{K\eta}(s')\,\sigma_{K\eta}(s')\,
    T^*_{K\eta,K\eta}(s')}{s'(s'-s-i\epsilon)} \ .
\label{mo_once}
\end{eqnarray}
The solution is obtained by the same method as for the twice subtracted
relations, with a modified inhomogeneity
\be
{\bf F}^\chi=\big(F_{K\pi}^\chi(0),\dots,
F_{K\pi}^\chi(0),F_{K\eta}^\chi(0),\dots,F_{K\eta}^\chi(0)\big)^T \ ,
\ee
[compare to Eq.~(\ref{inh})] and a modified kernel
\be
K_{\mu\nu}(s,s')=\frac{s}{\pi}\frac{\sigma_\nu(s')
T^*_{\nu\mu}(s')}{s'}\, \Theta(s'-s_{{\rm thres. }\nu})
\ee
[compare to Eq.~(\ref{kernel2})]. As Fig.~\ref{fig:comp_once} shows, the slope
of the chiral NLO expression is not matched any more. Also, the prediction of
the form factor at the Callan-Treiman point is wrong. As expected, the
uncertainty from the high-energy behavior is significantly larger for the
once-subtracted version (shaded areas). Moreover, the difference to the
twice-subtracted solution is larger than the uncertainties from the high-energy
input. 

Finally, we compare to the one-channel version of the form factor. For this, we
use the one-channel fit to the phase shift discussed in
Sec.~\ref{sec:scattering} and shown in Fig.~\ref{fig:phase} with the thick
dashed (blue) line. The twice-subtracted dispersion relation 
\be
F_{K\pi }(s)&=&F^\chi_{K\pi}(0)+(F^\chi_{K\pi})'(0)\,s
    +\frac{s^2}{\pi}\int\limits_{s_{K\pi }}^\infty ds'\,
    \frac{F_{K\pi }(s')\,\sigma_{K\pi }(s')\,
    T^*_{K\pi ,K\pi }(s')}{s'^2(s'-s-i\epsilon)} 
\label{dispone}
\ee
has the Omn\`es solution (cf., e.g., Ref.~\cite{Guerrero:1997ku})
\be
 F_{K\pi }(s)&=&P(s)\,F^\chi_{K\pi}(0)\,\exp
 \Big[s\,\frac{(F^\chi_{K\pi})'(0)}{F^\chi_{K\pi}(0)}
 +\frac{s^2}{\pi}\int\limits_{s_{K\pi}}^\infty 
 \frac{ds'}{s'^2}\,\frac{\delta(s')}{s'-s}\Big]
\label{omnes}
\ee
with a polynomial ambiguity $P(s)=1+a_2s^2+a_3s^3+\dots$ (note that
$\sigma\,T^*=\sin\delta\exp(-i\delta)$).

\begin{figure}\begin{center}
\includegraphics[scale=0.5]{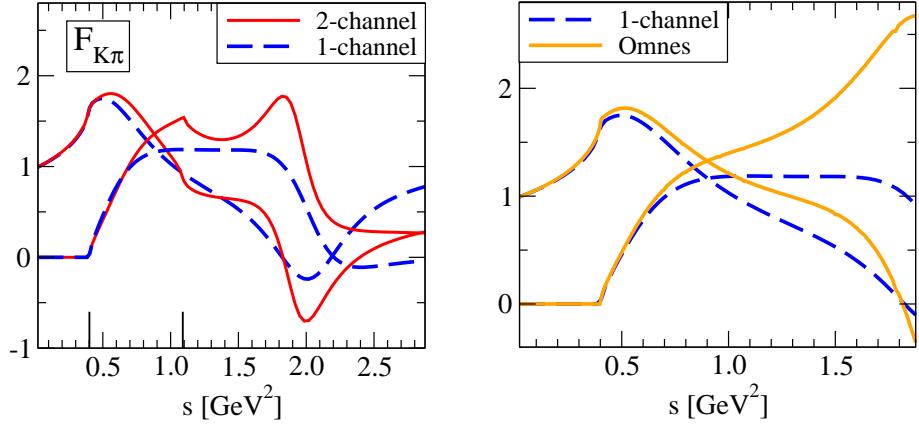}
\caption{Left: One-channel $K\pi$ form factor (dashed blue lines) compared to
the two-channel solution of Fig.~\ref{fig:result_FF} (solid red lines). Right:
Direct inversion of Eq.~(\ref{dispone}) (dashed blue lines) compared to the
Omn\`es solution according to Eq.~(\ref{omnes}). Both solutions are matched to
ChPT at the origin and differ from each other by a real polynomial  in $s$ of
degree 2 and higher, as must be.} 
\label{fig:onechannel}
\end{center}
\end{figure}
Using the numerical methods of Sec.~\ref{sec:solu}, Eq.~(\ref{dispone}) can be
directly inverted. To the left in Fig.~\ref{fig:onechannel}, the corresponding
result (dashed blue lines) is compared to the two-channel solution (solid red
lines). While the one-channel and the two-channel solutions are qualitatively
similar, at higher energies there are differences. In particular, the
two-channel solution never becomes zero, while the one-channel solution has a
zero around $s=2.2\,{\rm GeV}^2$. In Fig.~\ref{fig:onechannel} to the right, the
one-channel solution is compared to the Omn\`es solution of Eq.~(\ref{omnes})
with $P(s)=1$. The results start to differ already at quite low energies. The
Omn\`es solution cannot become zero while the direct inversion of
Eq.~(\ref{dispone}) exhibits a zero. As we have checked, the two solutions
indeed differ by a real polynomial $P(s)$ with a zero, of the form quoted below
Eq.~(\ref{omnes}). The role of zeroes in form factors has extensively been
discussed in Ref.~\cite{Oller:2007xd}. In summary, we observe that the
two-channel solution becomes small but stays non-zero, the one-channel solution
has a zero, and the Omn\`es solution cannot have a zero. Experimental data can
clarify whether a zero is present or not.


\subsection{Scalar $\pi\pi$ and $K\bar K$ form factors} 
\label{sec:pipiff}

In terms of  the isoscalar $S$-wave states  
\begin{eqnarray} 
&|\pi\pi\rangle_{\mathrm{I=0}}^{} & = \frac{1}{\sqrt{3}}
\left|\pi^+\pi^-\right\rangle + \frac{1}{\sqrt{6}}
\left|\pi^0\pi^0\right\rangle, \\ 
&|K\bar K\rangle_{\mathrm{I=0}}^{} & = 
\frac{1}{\sqrt{2}}\left|K^+K^-\right\rangle + 
\frac{1}{\sqrt{2}}\left|K^0\bar K^0\right\rangle,
\end{eqnarray}
the scalar form factors for the $\pi$ and $K$  mesons are defined as
\begin{eqnarray}
\sqrt{2}B_0\, F^{n/s}_1(s) &=& \langle 0|\bar nn /\bar ss|\pi\pi 
\rangle_{\mathrm{I=0}}^{}, 
\label{FFdef} \\
\sqrt{2}B_0\,F^{n/s}_2(s) &=& \langle 0|\bar nn/\bar ss|K\bar K 
\rangle_{\mathrm{I=0}}^{}, 
\nonumber  
\end{eqnarray}
where $\bar nn = (\bar uu+\bar dd)/\sqrt2$ denotes the non-strange scalar 
current, and  the notation ($\pi$ = 1, $K$ = 2) has been introduced  for
simplicity. Expressions  have already been derived in CHPT  up to
next-to-leading 
order~\cite{Gasser:1983yg,Gasser:1984gg,Gasser:1984ux,Meissner:2000bc}:
\begin{eqnarray}
&F_1^n(s)\:\: = & \left. \sqrt{\frac{3}{2}} \right[ 1 
+ \mu_\pi - \frac{\mu_\eta}{3}
+ \frac{16 m_\pi^2}{f^2}\left(2L_8^r-L_5^r\right)
+ 8\left(2L_6^r-L_4^r\right)\frac{2m_K^2 + 3m_\pi^2}{f^2}
+ \frac{8s}{f^2} L_4^r + \frac{4s}{f^2} L_5^r 
\nonumber \\ && + \left.
\left(\frac{2s - m_\pi^2}{2f^2}\right) J^r_{\pi\pi}(s)
+ \frac{s}{4f^2} J^r_{KK}(s)
+ \frac{m_\pi^2}{18f^2} J^r_{\eta\eta}(s)
\right],   \\
&F_1^s(s)\:\: = & \frac{\sqrt{3}}{\:2} \left[  
\frac{16 m_\pi^2}{f^2}\left(2L_6^r-L_4^r\right) + \frac{8s}{f^2} L_4^r
+ \frac{s}{2f^2} J^r_{KK}(s)
+ \frac{2}{9}\frac{m_\pi^2}{f^2} J^r_{\eta\eta}(s)
\right],\\
&F_2^n(s)\:\: = & \left. \frac{1}{\sqrt{2}} \right[ 1 
+ \left.\frac{8 L_4^r}{f^2}
\left(2s - m_\pi^2 - 6 m_K^2\right)
+ \frac{4 L_5^r}{f^2} 
\left(s - 4 m_K^2\right)
+ \frac{16 L_6^r}{f^2} \left(6 m_K^2 + m_\pi^2\right)
+ \frac{32 L_8^r}{f^2}\,m_K^2
+ \frac{2}{3} \mu_\eta
\right. \nonumber \\ && + \left.
\left(\frac{9s - 8 m_K^2}{36f^2}\right) J^r_{\eta\eta}(s)
+ \frac{3s}{4f^2} J^r_{KK}(s)
+ \frac{3s}{4f^2} J^r_{\pi\pi}(s)
\right], \\
&F_2^s(s)\:\: = & 1 
+ \frac{8 L_4^r}{f^2} \left(s - m_\pi^2 - 4 m_K^2\right)
+ \frac{4 L_5^r}{f^2} \left(s - 4 m_K^2\right)
+ \frac{16 L_6^r}{f^2} \left(4 m_K^2 + m_\pi^2\right)
+ \frac{32 L_8^r}{f^2}\,m_K^2
+ \frac{2}{3} \mu_\eta
\nonumber \\ && +
\left(\frac{9s - 8 m_K^2}{18f^2}\right) J^r_{\eta\eta}(s)
+ \frac{3s}{4f^2} J^r_{KK}(s). 
\end{eqnarray} 
Imposing  the unitarity constraints, the scalar form factor can be expressed  in terms of
the algebraic  coupled-channel equation
\begin{eqnarray}
F(s) &=& [I+K(s)\,g(s)]^{-1} R(s) \label{G_eq} \\
&=& [I-K(s)\,g(s)]\:R(s) \:+\: \mathcal{O}(p^6), \nonumber
\end{eqnarray}
where $R(s)$ has no right-hand cut and   in the second line, the equation
has been expanded up to NLO  in the  chiral expansion.   $K(s)$ is the $S$-wave  projected  kernel of
meson-meson scattering amplitudes that can be derived from the  leading-order chiral Lagrangian:
\begin{eqnarray}
&& K_{11} = \frac{2s - m_\pi^2}{2f^2}, \quad
   K_{12} = K_{21} = \frac{\sqrt{3}s}{4f^2},  \;\;\; 
   K_{22} = \frac{3s}{4f^2} \ .  
\nonumber 
\end{eqnarray} 
 The  loop integral can be  calculated either in the 
cutoff-regularization scheme with $q_{\rm max}\sim 1$GeV being the cutoff [cf. Erratum of Ref.~\cite{Oller:1998hw} for an explicit expression] or  in dimensional regularization with the modified
$\overline{\rm MS}$ subtraction scheme. In the latter scheme,  the meson loop
function $g_i(s)$ is  given by 
\begin{eqnarray}
J_{ii}^r(s) \!&\equiv&\! \frac{1}{16\pi^2}\left[
1 - \log\left(\frac{m_i^2}{\mu^2}\right) - \sigma_i(s)\log\left(
\frac{\sigma_i(s)+1}{\sigma_i(s)-1}\right)\right] \nonumber \\
&=& -g_i(s).
\label{m_loop} 
\end{eqnarray}
with  $
\sigma_i(s) = \sqrt{1- {4m_i^2}/{s}}$.
The matching between these two renormalization is given in Eq.~(\ref{matching}) with $q_{\rm max}=\Lambda$. 

\begin{figure}\begin{center}
\includegraphics[scale=0.6]{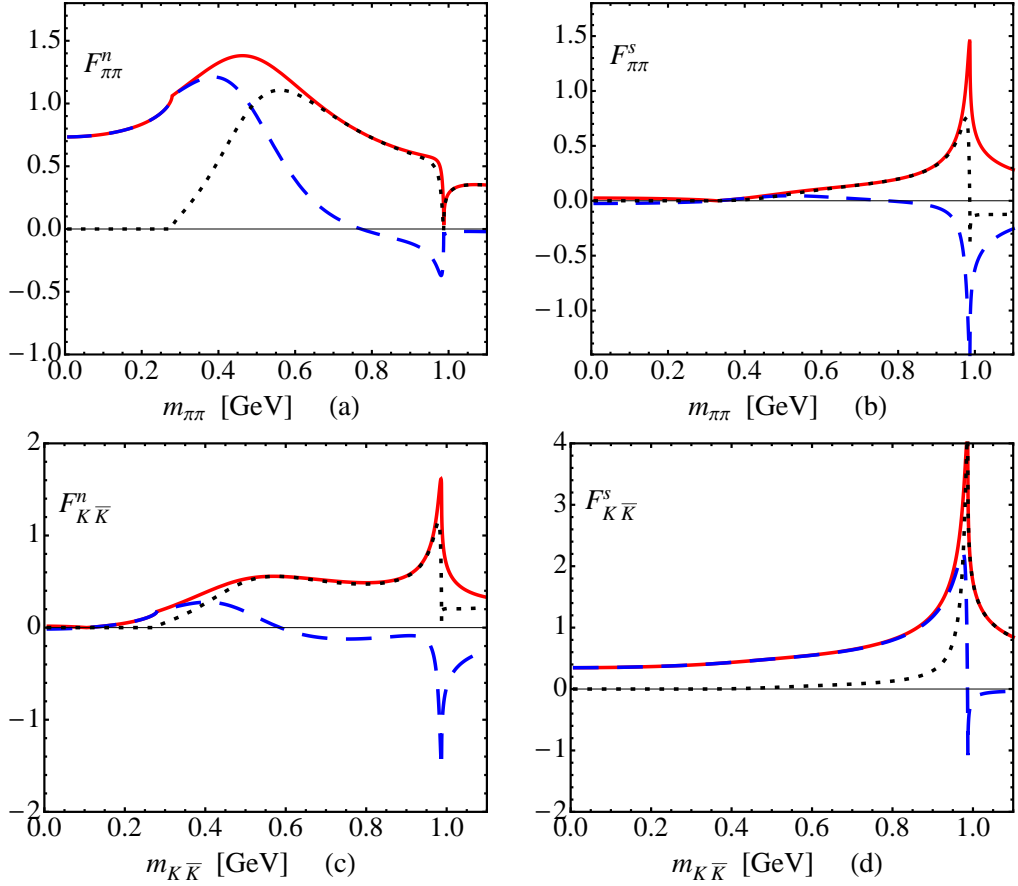}
\caption{The non-strange  and strange $\pi\pi$ and  $K\bar K$    scalar form factors  obtained in the unitarized chiral perturbation theory. The modulus, real part and imaginary part are shown in solid, dashed and dotted curves. } \label{fig:pipi_ff}
\end{center}
\end{figure}
The expressions for the $R^i$ are obtained  by matching the unitarization and
chiral perturbation theory~\cite{Oller:1997ti,Lahde:2006wr}:  
\begin{eqnarray}
&R_1^n(s)\:\: = & \sqrt{\frac{3}{2}} \bigg\{ 1 + \mu_\pi - 
\frac{\mu_\eta}{3}
+ \frac{16 m_\pi^2}{f^2}\left(2L_8^r-L_5^r\right)
+ 8\left(2L_6^r-L_4^r\right)\frac{2m_K^2 + 3m_\pi^2}{f^2}
+ \frac{8s}{f^2} L_4^r + \frac{4s}{f^2} L_5^r  \nonumber \\
&& -  \frac{m_\pi^2}{288\pi^2 f^2} \bigg[1 + 
\log\left(\frac{m_\eta^2}{\mu^2}\right)\bigg]
\bigg\}, \\
&R_1^s(s)\:\: = & \frac{\sqrt{3}}{\:2} \left\{  
\frac{16 m_\pi^2}{f^2}\left(2L_6^r-L_4^r\right)
+\frac{1}{3} \frac{8s}{f^2} L_4^r - \frac{m_\pi^2}{72\pi^2 f^2}
\left[1 + \log\left(\frac{m_\eta^2}{\mu^2}\right)\right]
\right\},\\
&R_2^n(s)\:\: = & \frac{1}{\sqrt{2}} \left\{
1 + \frac{8 L_4^r}{f^2} \left(2s - 6m_K^2 - m_\pi^2\right)
+ \frac{4 L_5^r}{f^2} \left(s - 4m_K^2\right)
+ \frac{16 L_6^r}{f^2} \left(6m_K^2 + m_\pi^2\right)
+ \frac{32 L_8^r}{f^2} m_K^2 + \frac{2}{3} \mu_\eta \right.\nonumber \\
&& + \left.\frac{m_K^2}{72\pi^2 f^2}
\left[1 + \log\left(\frac{m_\eta^2}{\mu^2}\right)\right]
\right\}, \\
&R_2^s(s)\:\: = & 1 
+ \frac{8 L_4^r}{f^2} \left(s - 4m_K^2 - m_\pi^2\right)
+ \frac{4 L_5^r}{f^2} \left(s - 4m_K^2\right)
+ \frac{16 L_6^r}{f^2} \left(4m_K^2 + m_\pi^2\right)
+ \frac{32 L_8^r}{f^2} m_K^2 + \frac{2}{3} \mu_\eta \nonumber \\
&& + \frac{m_K^2}{36\pi^2 f^2} 
\left[1 + \log\left(\frac{m_\eta^2}{\mu^2}\right)\right],
\end{eqnarray}
where the factor $1/3$ in $R_1^s(s)$ is missing in Ref.~\cite{Lahde:2006wr}.

With the above formulae and the fitted results for the low-energy constants
$L_i^r$ in Ref.~\cite{Lahde:2006wr} (evolved  from $M_\rho$ to the scale $\mu=
2q_{\rm max}/\sqrt{e}$), we show the non-strange and strange  $\pi\pi$ and
$K\bar K$ form factors in Fig.~\ref{fig:pipi_ff}.  The modulus, real part and
imaginary part are shown as solid, dashed and dotted curves. As the
figure shows, the chiral unitary ansatz predicts a form factor $F^n_{\pi\pi}$
with a zero close to the $\bar KK$ threshold. This feature has been extensively
discussed in Ref.~\cite{Oller:2007xd}.


\section{Results and Discussions }\label{sec:results}
In the narrow-width limit,  the integration over the $K\pi$ invariant mass leads
to the normalization of the $K^*$ [$K^*\equiv K^*(892)$] line shape: 
\begin{eqnarray}
 \int dm_{K\pi}^2 |L_{K^*}(m_{K\pi}^2)|^2= {\cal B}(\bar K^{*0}\to K^-\pi^+)
 \simeq \frac{2}{3}
\end{eqnarray}
with ${\cal B}$ being the branching ratio. Considering the  {momentum dependence
of the $K^*$ decay},  we have the running width as  
\begin{eqnarray}
 \Gamma_{K^*} (m_{K\pi}^2) =  \Gamma_{K^*}^0 \left(\frac{ |\vec q\,|}{ |
 \vec q_0|}\right)^3   \frac{m_{K^*}}{m_{K\pi}}   
 \frac{1+ (R|\vec q_0|)^2}{1+ (R|\vec q\,|)^2},
\end{eqnarray}
and the Blatt-Weisskopf parameter $R=(2.1\pm 0.5\pm 0.5) {\rm
GeV}^{-1}$~\cite{delAmoSanchez:2010fd}.  It is plausible to assume the same form
for the $\phi$ meson except that the mass, total decay width and the branching
ratios into $K^+K^-$ are replaced correspondingly.

\begin{figure}\begin{center}
\includegraphics[scale=0.5]{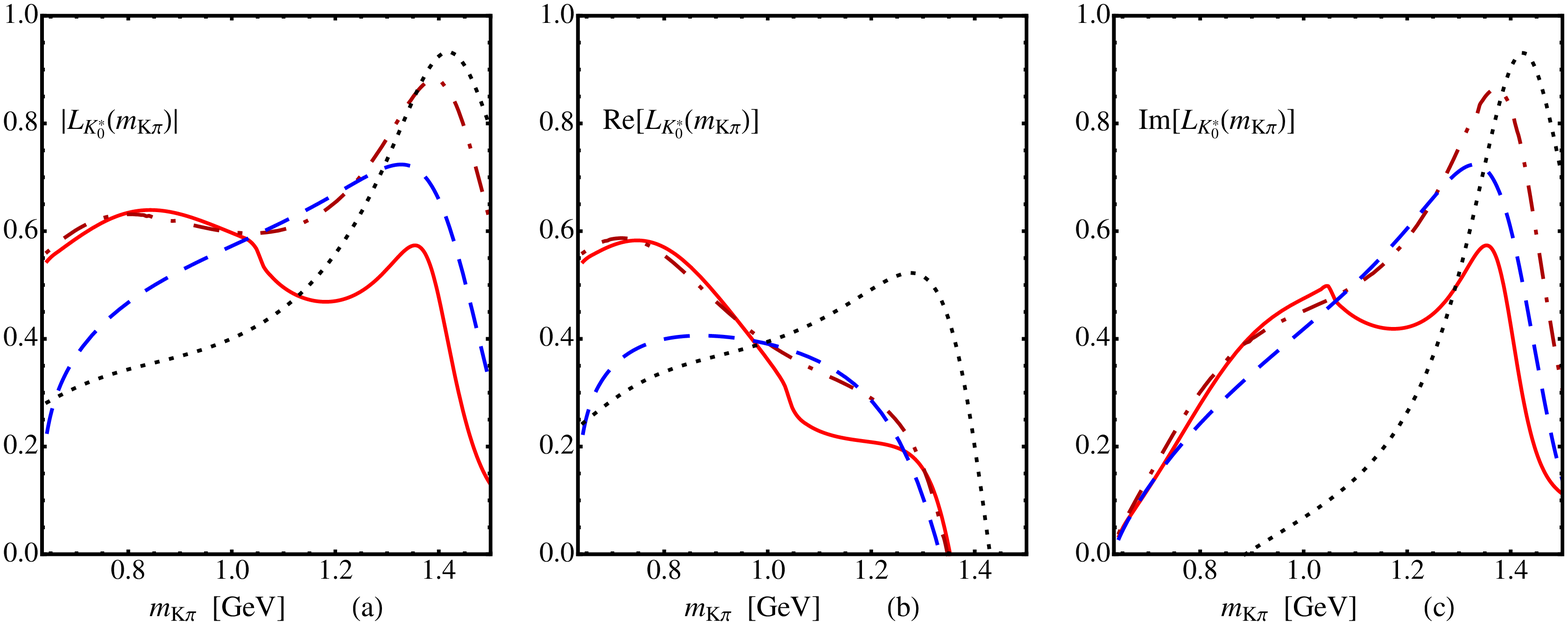}
\caption{S-wave $K\pi$ line shapes. The modulus, real part, and imaginary part
are given in panel (a), (b), and (c), respectively. Solid (red) lines: lineshape
given by Eq.~\eqref{eq:chiPTlineshape} with the Muskhelishvili-Omn\`es solution 
of Eq.~(\ref{mo}) for the form factor.  Dash-dotted (dark red) lines: Omn\`es
solution of Eq.~(\ref{omnes}). Dotted lines: LASS lineshape of
Eq.~\eqref{eq:LASS}; dashed (blue) lines: parameterization of  Eq.~\eqref{eq:BT}.
}
\label{fig:lineshapecomparison}
\end{center}
\end{figure}

We start  with the numerical  discussion on the $K^-\pi^+$ and $K^+K^-$
line-shapes. In the case of $K^-\pi^+$, we compare with the data-inspired  LASS
parametrization~\cite{Aston:1987ir,Aubert:2005ce}
\begin{eqnarray}
 L_{K^*_0}^{\rm LASS}(m_{K\pi}) =   {\cal N}_{\rm LASS} 
 \left( \frac{m_{K\pi}}{|\vec q\,|(\cot \delta_B -i)}  + e^{2i\delta_B}  
 \frac{m_{K^*_0} ^2\Gamma_{K^*_0}^0 /|\vec q_0|   }{  m_{K^*_0}^2- m_{K\pi}^2 
 -i  m_{K^*_0} \Gamma_{K^*_0} (m_{K\pi}^2) } \right),  
\label{eq:LASS}
\end{eqnarray}
with $\vec q$ and $\vec q_0$ being three momentum  of kaon/pion at $m_{K\pi}$ and $m_{K^*_0}$, and the parameters~\cite{Aubert:2008aa}
\begin{eqnarray}
 \cot\delta_B = \frac{1}{a|\vec q\,|} + \frac{1}{2} r|\vec q\,|, \;\; 
  a= 1.94\, {\rm GeV}^{-1}, \;\; r= 1.76\, {\rm GeV}^{-1}. 
\end{eqnarray}
The running decay width in Eq.~\eqref{eq:LASS} is 
\begin{eqnarray}
 \Gamma_{K^*_0} (m_{K\pi}^2) =  \Gamma_{K^*_0}^0  \frac{ |\vec q\,|}{ 
 |\vec q_0|}    \frac{m_{K^*_0}}{m_{K\pi}}. 
\end{eqnarray}
The constant ${\cal N}_{\rm LASS}$ in Eq.~\eqref{eq:LASS}  can be chosen such
that the  $K^*_0(1430)$ term has the same normalization as the ordinary
Breit-Wigner formula.  In the study of S-wave effects in $B\to K^* l^+l^-$,  
the authors of  Ref.~\cite{Becirevic:2012dp} have suggested a phenomenological
parametrization: 
\begin{eqnarray}
 L_{K^*_0}^{\rm BT} = {\cal N}_{\rm BT} \left[ \frac{g_{\kappa}}{ m_{K\pi}^2
 - (m_{\kappa}- i\Gamma_{\kappa}/2)^2} -\frac{1}{ m_{K\pi}^2- (m_{K^*_0(1430)}
 - i\Gamma_{K^*_0(1430)}/2)^2} \right], \label{eq:BT}
\end{eqnarray}
with the coupling constant chosen by hand $g_{\kappa}\sim 0.2$ and ${\cal
N}_{BT}$ as the normalization constant. 

In Fig~\ref{fig:lineshapecomparison}, we compare the three parametrizations
defined in Eq.~\eqref{eq:chiPTlineshape},  Eq.~\eqref{eq:LASS}, and
Eq.~\eqref{eq:BT} corresponding to the solid, dotted and dashed curves  {}{with the
result of Eq.~(\ref{eq:chiPTlineshape}) given by the Muskhelishvili-Omn\`es
solution according to Eq.~(\ref{mo}) and shown in Fig.~\ref{fig:result_FF}.}  Dash-dotted  lines are the  Omn\`es
solution of Eq.~(\ref{omnes}).   In the experimental study of $B\to
J/\psi\, K\pi$~\cite{Aubert:2008aa},  the Babar collaboration found in their
Fig.11 that the LASS parametrization for S-wave contribution will undershoot the
experimental data in the low $K\pi$ invariant mass region, in particular around
$m_{K\pi}\sim 0.7$ GeV.  As Fig.~\ref{fig:lineshapecomparison} shows for $|L_{K^*_0}(m_{K\pi})|$, the results derived in this study can indeed improve this underprediction at low invariant masses. Both the Muskhelishvili-Omn\`es and the Omn\`es solution have more strength at low invariant masses than the LASS parametrization. In both results of this study, a double hump structure from the $\kappa$ and the $K_0^*(1430)$ is observed. In the two-channel solution (solid red line), the flux into the $K\eta$ channel significantly reduces the strength at the $K_0^*(1430)$ position, which is not the case for the one-channel (elastic) Omn\`es solution (dash-dotted lines). As discussed before, the latter cannot have a zero and thus might overpredict the strength at the $K_0^*(1430)$ position.
 
We will also study the process $B_s\to \phi(\to K^+K^-) l^+l^-$,  where the
$K\bar K$ state is  close to threshold and  the S-wave mass distribution {is}
well-described by the Flatt\'e model~\cite{Flatte:1976xu}
\begin{eqnarray}
 L_{f_0(980)}(m_{K\bar K}) = \sqrt{\frac{ m_{f_0}  \Gamma_{f_0\to K^+K^-} }{\pi} }
 \frac{1}{ m_{K\bar K}^2-  m_{f_0}^2 +i  m_{K\bar K}(g_1 \rho_{\pi\pi}+ g_2 \rho_{K\bar K}) },
\end{eqnarray}
with $\rho_{\pi\pi /K\bar K}$ being the Lorentz-invariant phase space, $\rho_{\pi\pi /K\bar K}=2|\vec
q\,|/m_{K\bar  K}= \sqrt{1- 4m_{\pi/K}^2/m_{K\bar K}^2}$.   The involved  couplings have
been measured from $J/\psi$ decays by  {}{the} BES
collaboration~\cite{Ablikim:2004wn}
\begin{eqnarray}
 m_{f_0}  \Gamma_{f_0\to K^+K^-} =\frac{1}{2} g_2 \rho_{K\bar K},\;\; g_1 
 = (165 \pm 10\pm 15){\rm MeV},\;\; g_2/g_1= 4.21\pm 0.25\pm 0.21. 
\end{eqnarray} 

\begin{figure}\begin{center}
\includegraphics[scale=0.5]{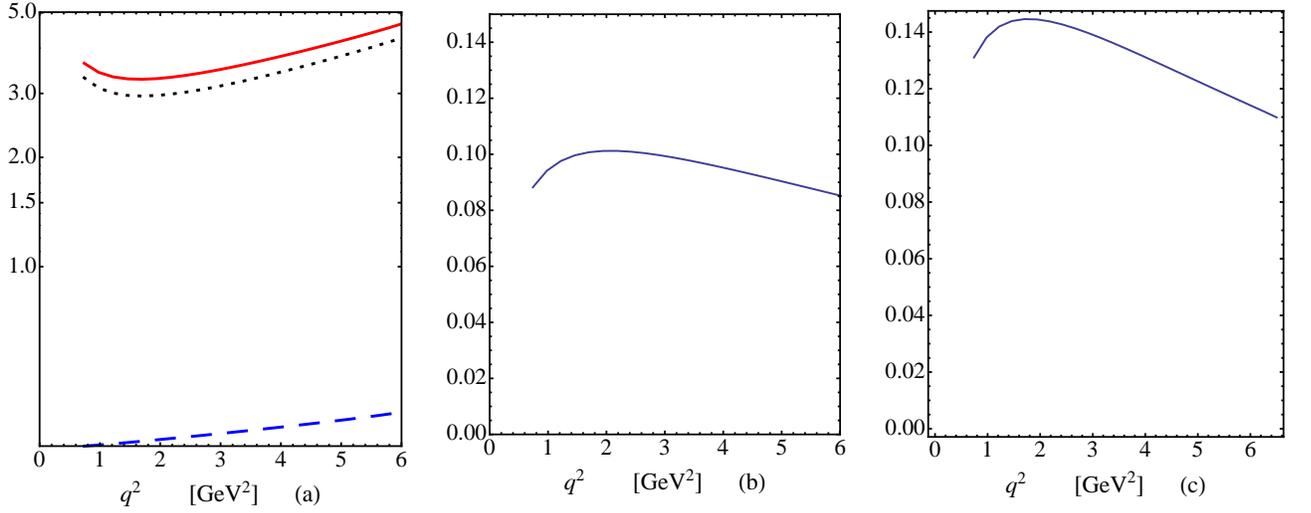} 
\caption{The S-wave contribution and its interference with P-wave to the
differential decay width in $\bar B^0\to K^-\pi^+ l^+l^-$. In panel (a), the
dashed, dotted and solid curves  denote the S-wave, P-wave and  total
contributions $d\Gamma/dq^2$ in units of $10^{-8} {\rm GeV}^{-2}$. Panel (b)
shows the S-wave fraction $d\Gamma_S/dq^2$. We also show the forward-back
asymmetry $dA_{FB}^{K}/dq^2$ for the charged kaon in panel (c).
} 
\label{fig:dgammadq2_Kpi}
\end{center}
\end{figure}

\begin{figure}\begin{center}
\includegraphics[scale=0.5]{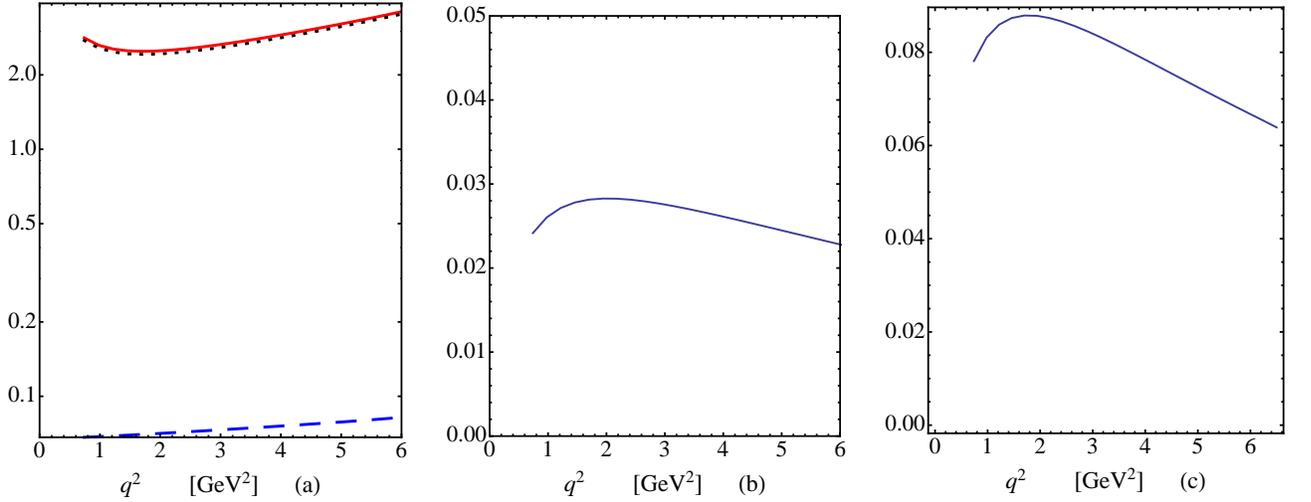}
\caption{ Same as Fig.~\ref{fig:dgammadq2_Kpi} but for $\bar B_s^0\to K^+K^-l^+l^-$.  } \label{fig:dgammadq2_KKbar}
\end{center}
\end{figure}

To quantitatively  demonstrate the size of S-wave contributions, we study the
total  differential decay width $d\Gamma/dq^2$ (integrated over $m_{K\pi}$),
the  S-wave fraction and  the    forward-backward asymmetry distribution  for
the charged kaon, ${dA_{FB}^{K}}/{ dq^2 } $: 
\begin{eqnarray}
 \frac{d\Gamma}{ dq^2}
 &\equiv&  \int_{(m_{K^*}- \delta m)^2} ^{(m_{K^*}
 + \delta m)^2}  dm_{K\pi}^2\frac{d^2\Gamma}{ dq^2 dm_{K\pi}^2 },
 \nonumber\\
 \frac{d\Gamma_S}{ dq^2}
 &\equiv&  \int_{(m_{K^*}- \delta m)^2} ^{(m_{K^*}+ \delta m)^2}  dm_{K\pi}^2
 \frac{d^2\Gamma_S}{ dq^2 dm_{K\pi}^2 } 
 = \int_{(m_{K^*}- \delta m)^2} ^{(m_{K^*}+ \delta m)^2}  
 dm_{K\pi}^2[|A_{L0}|^2+ |A_{R0}|^2],
 \nonumber\\
 \frac{dA_{FB}^{K}}{ dq^2 } 
 &\equiv&  \int_{(m_{K^*}- \delta m)^2} ^{(m_{K^*}+ \delta m)^2}  
 dm_{K\pi}^2\frac{d^2A_{FB}^K}{ dq^2 dm_{K\pi}^2 },\;\;\; \;\;\; \overline{
 \frac{dA_{FB}^{K}}{ dq^2 }} \equiv  \frac{
 \frac{dA_{FB}^{K}}{ dq^2 } }{ \frac{d\Gamma}{ dq^2}},
\end{eqnarray} 
with $m_{K^*}\equiv m_{K^*(892)}$. 

With the   choice of  $\delta_{m} = 100$ MeV (the default choice adopted by the
LHCb collaboration~\cite{Aaij:2013iag}),   we show our results for $d\Gamma/dq^2$  
in panel (a) of  Fig.~\ref{fig:dgammadq2_Kpi},  where the dashed,
dotted and  solid curves  denote the S-wave, P-wave and  total  contributions
respectively. The panel (b) and (c) correspond to  the S-wave fraction and  and
the forward-backward asymmetry. In the case of $B_s\to \phi l^+l^-$ the bin size
is chosen as $\delta_{m} = 20$ MeV and the corresponding results are shown in
Fig.~\ref{fig:dgammadq2_KKbar}.     From these figures, we find that the S-wave
contribution can
reach 10\% in $\bar B^0\to K^-\pi^+ l^+l^-$, while it is  about  5\% in $B_s\to K^+K^-l^+l^-$.    It is necessary to stress that there is a
sign ambiguity  in the forward-backward asymmetry $ {dA_{FB}^{K}}/{ dq^2
}$ from the use of Watson theorem.  
However, the magnitude  shown in panel (c) is sizable and  thus measurable  in
future. Since this quantity $ {dA_{FB}^{K}}/{ dq^2 }$ arises from the S-wave
and P-wave interference, it  can be used to constrain the S-wave meson-meson
scattering when precise data is available in future. 

Based on the data sample of $1fb^{-1}$,  the LHCb collaboration has set an upper
limit for the integrated S-wave fraction $F_S< 0.07$ at $68\%$ CL, in the range $1\,{\rm
GeV}^2 < q^2<6\, {\rm GeV}^2$~\cite{Aaij:2013iag} and  
\begin{eqnarray}
F_S =0.04\pm 0.04. 
\end{eqnarray}
Our estimate is slightly larger than  {this} but still consistent with the data
when   errors are taken into account. 

\begin{figure}\begin{center}
\includegraphics[scale=0.6]{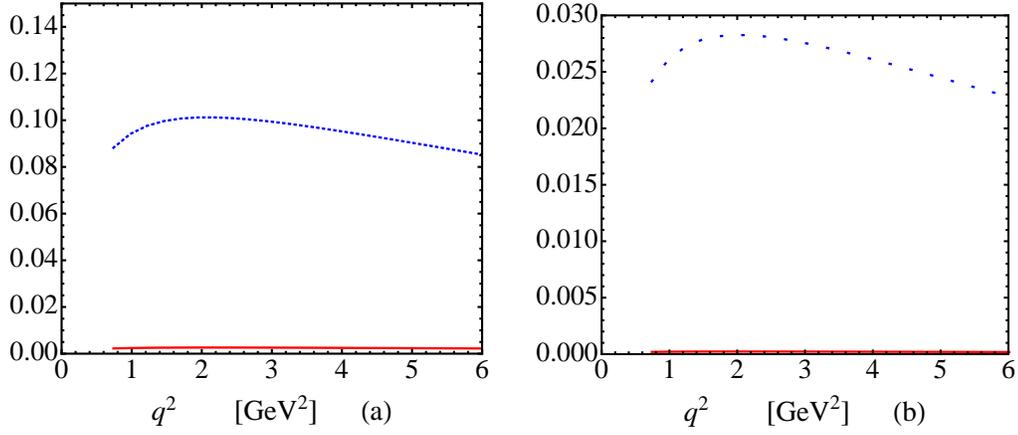}
\caption{Fraction of S-wave contributions shown in the dotted (blue) curves in
$\bar B^0\to K^-\pi^+l^+l^-$ (left panel) and $\bar B_s\to K^+K^-l^+l^-$ (right
panel). After using the kinematical subtraction in Eq.~\eqref{eq:substraction}, 
the S-wave contributions are  reduced to less than 1\%.    }
\label{fig:SwaveSub}
\end{center}
\end{figure}

To minimize the S-wave contributions and project out the P-wave, we  propose to
study the subtracted differential decay width:
\begin{eqnarray}
 \frac{d\Gamma'}{ dq^2}
 &\equiv&  \left[ 2\int_{(m_{K^*}- \delta m_2)^2} ^{(m_{K^*}+ \delta m_2)^2}  
 -\int_{(m_{K^*}- \delta m_1)^2} ^{(m_{K^*}+ \delta m_1)^2}\right] 
 dm_{K\pi}^2\frac{d^2\Gamma}{ dq^2 dm_{K\pi}^2 } \nonumber\\
 &=&\left[  \int_{(m_{K^*}- \delta m_2)^2} ^{(m_{K^*}+ \delta m_2)^2}  
 -\int_{(m_{K^*}- \delta m_1)^2} ^{(m_{K^*}- \delta m_2)^2}
 -\int_{(m_{K^*}+\delta m_2)^2} ^{(m_{K^*}+ \delta m_1)^2}\right] 
 dm_{K\pi}^2\frac{d^2\Gamma}{ dq^2 dm_{K\pi}^2 },
 \label{eq:substraction}
\end{eqnarray}
with $\delta m_1= 2\delta m_2$.  The physical interpretation  is to  slice the
$K\pi$ invariant mass distribution  into 4  bins around $m_{K\pi}\sim
m_{K^*(892)}$ with $\delta m_2$ being the  size of the  bins. We select the events
in the central two bins and subtract those from the other two bins.  For
illustration we use  $\delta_{m_1} =2\delta_{m_2} = 100$ MeV for $B\to
K^-\pi^+l^+l^-$ and  $\delta_{m_1} =2\delta_{m_2}= 20$ MeV   for  $B_s\to
K^+K^-l^+l^-$~\footnote{ The CDF collaboration has used the choice of
$\delta_{m} =50 {\rm MeV}$ and  $\delta_{m} =10 {\rm MeV}$ for $K^+\pi^-$ and
$K^+K^-$ states in the analysis of $B\to K^*l^+l^-$ and $B_s\to \phi l^+l^-$
decays~\cite{Aaltonen:2011cn}, while LHCb adopted  $\delta_{m} =100 {\rm MeV}$
for $K\pi$~\cite{Aaij:2013iag}, and  $\delta_{m} =12 {\rm MeV}$ for
$K^+K^-$~\cite{Aaij:2013aln}. }.  From the results in Fig.~\ref{fig:SwaveSub}, 
we can see the S-wave contributions have been reduced to less than 1\% for both
$B\to K^-\pi^+ l^+l^-$ and $B_s\to K^+K^-l^+l^-$.  The disadvantage is the demand  of larger statistics.

Before closing this section, several remarks  are given in order.  

\begin{itemize}
\item  
We have considered the invariant mass distribution for the light meson pair
using  chiral perturbation theory {and dispersion relations}, but  used the
heavy-to-light form factors calculated in the resonance approximation. 

\item 
Our analysis will be improved by a direct calculation of  the matrix element
$\langle (K\pi)_0|\bar s \Gamma b|\bar B\rangle$ and also hard-scattering QCD
corrections. This can be achieved in the factorization approach  by merging the
perturbative nature of B decays and the chiral perturbation description of
$K\pi$. For future use,  we parametrize these matrix elements  in the following
\begin{eqnarray}
 \langle (K\pi)_0(p_{K\pi})|\bar s \gamma_\mu\gamma_5 b|\overline B (p_B)
 \rangle  &=& -i  \frac{1}{m_{K\pi}} \bigg\{ \bigg[P_{\mu}
 -\frac{m_B^2-m_{K\pi}^2}{q^2} q_\mu \bigg] {\cal F}_{1}(m_{K\pi}^2, q^2) 
 +\frac{m_B^2-m_{K\pi}^2}{q^2} q_\mu  {\cal F}_{0}(m_{K\pi}^2, q^2)  \bigg\}, 
 \nonumber\\
 \langle (K\pi)_0(p_{K\pi})|\bar s \sigma_{\mu\nu} q^\nu \gamma_5 b|
 \overline B (p_B)\rangle  &=& \frac{{\cal F}_T(m_{K\pi}^2, 
 q^2)}{m_{K\pi}(m_B+m_{K\pi})} \bigg[ ({m_B^2-m_{K\pi}^2}) q_\mu - q^2 
 P_{\mu}\bigg],
 \label{eq:generalized_form_factors}
\end{eqnarray}
where ${\cal F}_{1,0,T}(m_{K\pi},q^2)$ are the  ``generalized'' form factors and
an additional dependence on kaon/pion momentum is suppressed here. Further, $P=
p_{B}+p_{K\pi}$ and $q=p_B-p_{K\pi}$.  Fortunately, the heavy quark and large
recoil symmetries are valid in the small $q^2$ region, and thus we have the
large-recoil symmetry at the leading-order in $\alpha_s$ and
$1/m_b$~\cite{Charles:1998dr}:
\begin{eqnarray}
 {\cal F}_1 (m_{K\pi},q^2)= \frac{m_{B}^2-m_{K\pi}^2}{m_{B}^2-m_{K\pi}^2-q^2 } 
 {\cal F}_0 (m_{K\pi},q^2)=  \frac{m_B}{m_B+m_{K\pi}}  {\cal F}_T 
 (m_{K\pi},q^2)\equiv {\cal \xi}(m_{K\pi},q^2). 
\end{eqnarray}

\item 
The calculation of  generalized form  factors ${\cal F}_{1,0,T}(m_{K\pi},q^2)$
and hard-scattering QCD corrections  will remove the necessity for the use of
Watson theorem,  but  requires the knowledge of the $K\pi$ generalized
light-cone distribution amplitudes, defined in the following form for the $K\pi$
system with spin $J$:
\begin{eqnarray}
 \langle (K\pi)_J|\bar s(z) \Gamma q(-z)|0\rangle.
\end{eqnarray}   
Here $\Gamma$ denotes a generic Dirac matrix.  It is interesting to notice these
distribution amplitudes are normalized to   scalar form factors which were
already discussed in the last section. In  the $\pi\pi$ case, the leading-twist 
distribution amplitude  has been derived using CHPT~\cite{Diehl:2005rn}, while
to the best of our knowledge the sub-leading twist distribution amplitudes are
not  yet available from first-principles in the literature.   

\item  
Our analysis can be generalized to  the  $b\to ul\bar \nu$ processes like
$B^-\to \pi^+\pi^- l\bar\nu$ and $\bar B_s^0\to K^0\pi^+ l\bar \nu$, which are
of great interest  for the determination of CKM matrix element $|V_{ub}|$. 

\item  
Similarly, if two final  mesons are moving collinear (with small invariant mass) in three-body $B$ decays
like $B\to KK\bar K, K\pi\pi, K\bar K\pi, \pi\pi\pi$,  the decay matrix element
will be factorized at leading order in $1/m_b$. Depending on the topology of
the two-meson system,  decay  amplitudes are expressed in terms of a product of the
generalized form factors for the recoiling  two-meson system as defined in
Eq.~\eqref{eq:generalized_form_factors} and the  light-cone distribution
amplitude for the emitted meson, or the heavy-to-light form factors for the
recoil meson and generalized distribution amplitude for the emitted two-meson system.  See for example Refs.~\cite{Chen:2002th,Cheng:2005ug,ElBennich:2009da} for some  discussions along this line. 
\end{itemize}

\section{Conclusion}

Our understanding of standard model and CP violation  benefits a lot from  
heavy flavor physics, and thereby considerable amount of effort has been made in
recent years. Since the momentum transfer in $B$ decays is  large, one can use
the factorization scheme to separate the short-distance  and the
long-distance physics. The calculation of the short-distance part is based on
perturbation theory and operator product expansion in QCD, and has reached a
high precision.  The long-distance  matrix element  is usually  challenged by
our knowledge of the S-wave.  In this work, we point out the S-wave contribution
can be controlled using chiral perturbation theory, which offers a systematic
way for the study of the S-wave in $B$ semi-leptonic and non-leptonic decays. 
Still, direct measurements have to be made to disentangle remaining polynomial
ambiguities.

In this work, we considered  the example  $B\to K\pi l^+l^-$ and identified the
matrix element  $\langle  K\pi|\bar s\Gamma b|\bar B\rangle$ with the S-wave
scalar form factor.  Using the S-wave $K \pi$ and $K \bar K$ form factors from 
{the Muskhelishvili-Omn\`es solution and} unitarized chiral perturbation
theory,  {respectively,}  we have investigated the S-wave  contributions in 
{the decay} $\overline B^0\to  K^- \pi^+ l^+l^-$ with the $K\pi$  invariant mass
lying in the vicinity of the mass of  {the} $K^*(892)$ and the {decay}
$B_s\to    K^- K^+ l^+l^-$ with $m_{KK}\sim m_{\phi}$.  We found that
differential decay widths are affected  by about 10\% in the process of
$\overline B^0\to  K^- \pi^+ l^+l^-$, which is larger  than but still consistent
with the LHCb measurements.   A forward-backward asymmetry  for the charged kaon
in the final state arises due to the interference between S-wave and P-wave
contributions.  The measurement of this asymmetry in the future offers  a new
way to constrain the variation of the  $K\pi$ S-wave phase versus the invariant 
mass  and it should be compared with other experimental determinations. 

\section*{Acknowledgements}

We are grateful to  Feng-Kun Guo, C. Hanhart, Bastian Kubis and J.A. Oller for 
enlightening discussions. W.W. thanks  Ignacio Bediaga and Manoel Robilotta  for
valuable discussions during the FPCP2013 conference.  This work is supported in
part by the DFG and the NSFC through funds provided to the Sino-German CRC 110
``Symmetries and the Emergence of Structure in QCD'', the ``EU I3HP Study of
Strongly Interacting Matter''  under the Seventh Framework Program of the EU.

\begin{appendix}


\section{$B\to K^*_J$ Form factors}
 
The $B\to K_0^*$ form factors are parametrized as
\begin{eqnarray}
    \langle K^*_0(P_2) |\bar s  \gamma_\mu\gamma_5  b|
    \overline  {B}(P_{B})\rangle
    &=&-i\left\{\left
    [P_\mu - \frac{m_{B}^2-m_{K^*_0}^2}{q^2}q_\mu \right ]
    F_1 (q^2)
    +\frac{m_{B}^2-m_{K^*_0}^2}{q^2}q_\mu F_0 (q^2)\right\} ,\nonumber\\
   \langle  K^*_0(P_2) |\bar s  \sigma_{\mu\nu} q^\nu\gamma_5  b|
   \overline  {B}(P_{B})\rangle &=&\left[ (m_{B}^2-m_{K^*_0}^2) q_\mu 
   - q^2 P_\mu\right ]
  \frac{F_T (q^2)}{
   m_{B}+m_{K^*_0}},
   \end{eqnarray}
while the  $B\to K_J^*(J \geqslant 1)$ form factors  are defined
by~\cite{Hatanaka:2009sj,Wang:2010ni,Yang:2010qd}
 \begin{eqnarray}
  \langle K_J^*(P_2,\epsilon)|\bar s\gamma^{\mu}b|\overline B(P_B)\rangle
   &=&-\frac{2V(q^2)}{m_B+m_{K_J^*}}\epsilon^{\mu\nu\rho\sigma} 
   \epsilon^*_{J\nu}  P_{B\rho}P_{2\sigma}, \nonumber\\
  \langle  K_J^*(P_2,\epsilon)|\bar s\gamma^{\mu}\gamma_5 b|\overline
  B(P_B)\rangle
   &=&2im_{K_J^*} A_0(q^2)\frac{\epsilon^*_{J } \cdot  q }{ q^2}q^{\mu}
    +i(m_B+m_{K_J^*})A_1(q^2)\left[ \epsilon^*_{J\mu }
    -\frac{\epsilon^*_{J } \cdot  q }{q^2}q^{\mu} \right] \nonumber\\
    &&-iA_2(q^2)\frac{\epsilon^*_{J} \cdot  q }{  m_B+m_{K_J^*} }
     \left[ P^{\mu}-\frac{m_B^2-m_{K_J^*}^2}{q^2}q^{\mu} \right],\nonumber\\
  \langle  K_J^*(P_2,\epsilon)|\bar s\sigma^{\mu\nu}q_{\nu}b|\overline
  B(P_B)\rangle
   &=&-2iT_1(q^2)\epsilon^{\mu\nu\rho\sigma} \epsilon^*_{J\nu} 
   P_{B\rho}P_{2\sigma}, \nonumber\\
  \langle  K_J^*(P_2,\epsilon)|\bar s\sigma^{\mu\nu}\gamma_5q_{\nu}b|
  \overline  B(P_B)\rangle
   &=&T_2(q^2)\left[(m_B^2-m_{K_J^*}^2) \epsilon^*_{J\mu }
       - {\epsilon^*_{J } \cdot  q }  P^{\mu} \right] 
       +T_3(q^2) {\epsilon^*_{J } \cdot  q }\left[
       q^{\mu}-\frac{q^2}{m_B^2-m_{K_J^*}^2}P^{\mu}\right],
       \nonumber
\label{eq:BtoTformfactors-definition}
 \end{eqnarray}
in which we have adopted $\epsilon^{0123}=+1$,   $q=P_B-P_2$, and $P=P_B+P_2$.  
The polarization vector  $\epsilon_J$ is constructed by the rank-$J$
polarization tensor
\begin{eqnarray}
  &&\epsilon_{J\mu}(h) =\frac{1}{m_B^{J-1}}
  \epsilon_{\mu\nu_1 \nu_2 
  ...\nu_{J-1}}(h)P_{B}^{\nu_1}P_{B}^{\nu_2}...P_{B}^{\nu_{J-1}},
\end{eqnarray}
with  the helicity $h=0,\pm1$.  In the case of $J=1$, it  is reduced to the
ordinary polarization vector. 

These transition  form factors have been calculated  in the perturbative QCD
approach~\cite{Li:2008tk,Wang:2006ria,Shen:2006ms,Kim:2009dg} using the inputs from Ref.~\cite{Cheng:2005nb} and the results
are  summarized in Tab.~\ref{tab:resultsscenario1}~\footnote{Results for
heavy-to-light form factors have been updated using a  
package that can compute form factors and two-body nonleptonic $B$ decays in the perturbative
QCD approach available at:  http://www.itkp.uni-bonn.de/$\sim$weiwang/. }, in
which the dipole parametrization  has been adopted to access the momentum
transfer dependence 
\begin{eqnarray}
 F(q^2)= \frac{F(0)}{ 1+a_F q^2/m_B^2+ b_F (q^2/m_B^2)^2}. 
\end{eqnarray}
Light-cone QCD sum rules
results~\cite{Ball:2004rg,Khodjamirian:2006st,Colangelo:2010bg,Colangelo:2010wg} have similar
modulus and mass distributions for the transition  form factors, and thus our
discussion on semileptonic $B$ decays will  not be  affected. 

 \begin{table}
 \caption{$B\to K^*_J$ and $B_s\to f_0(980),\phi$ form factors in the
perturbative QCD approach.}
\label{tab:resultsscenario1}
 \begin{center}
 \begin{tabular}{|l c c l | l c c l|}
 \hline 
 \ \ \        &$F(0)$\hspace*{0.3cm} & $a_F$ \hspace*{0.3cm} &$b_F$ \hspace*{0.3cm} & & $F(0)$ \hspace*{0.3cm} & $a_F$ \hspace*{0.3cm} &$b_F$   \bigstrut\\
 \hline  
   $V^{\bar B^0{\bar K^{*0}}}$ \hspace*{0.9cm}& $0.25$&$-2.2$&$1.3$
 &$A_0^{\bar B^0{\bar K^{*0}}}$ & $0.29$&$-2.2$&$1.2$  \bigstrut[t]\\
 $A_1^{\bar B^0{\bar K^{*0}}}$ & $0.19$&$-1.3$&$0.16$
 &$A_2^{\bar B^0{\bar K^{*0}}}$ & $0.29$&$-2.2$&$1.2$  \\
 $T_1^{\bar B^0{\bar K^{*0}}}$ & $0.23$&$-2.2$&$1.2$
 &$T_2^{\bar B^0{\bar K^{*0}}}$ & $0.23$&$-1.2$&$0.068$  \\
 $T_3^{\bar B^0{\bar K^{*0}}}$ & $0.16$&$-1.7$&$0.98$  &&&&\\
 $F_1^{\bar B^0{\bar K_0^{*0}(800)}}$ & $0.27$&$-2.1$&$1.2$
 &$F_0^{\bar B^0{\bar K_0^{*0}(800)}}$ & $0.27$&$-1.2$&$0.080$  \\
 $F_T^{\bar B^0{\bar K_0^{*0}(800)}}$ & $0.30$&$-2.2$&$1.2$  &&&&\\ 
 $V^{\bar B_s\phi}$ & $0.26$&$-2.2$&$1.3$
 &$A_0^{\bar B_s\phi}$ & $0.30$&$-2.2$&$1.2$  \\
 $A_1^{\bar B_s\phi}$ & $0.19$&$-1.2$&$0.15$
 &$A_2^{\bar B_s\phi}$ & $0.30$&$-2.2$&$1.2$  \\
 $T_1^{\bar B_s\phi}$ & $0.23$&$-2.1$&$1.2$
 &$T_2^{\bar B_s\phi}$ & $0.23$&$-1.2$&$0.066$  \\
 $T_3^{\bar B_s\phi}$ & $0.15$&$-1.5$&$1.1$ &&&&\\
  $F_1^{\bar B_s{f_0(980)}}$ & $0.34$&$-2.1$&$1.1$
 &$F_0^{\bar B_s{f_0(980)}}$ & $0.34$&$-1.2$&$0.078$  \\
 $F_T^{\bar B_s{f_0(980)}}$ & $0.38$&$-2.1$&$1.2$ &&&&\bigstrut[b]\\
 \hline
 \end{tabular}
 \end{center}
 \end{table}

\section{Helicity amplitudes}

The differential distributions can be  expressed in terms of the helicity
amplitude (see for instance Ref.~\cite{Lu:2011jm}):
\begin{eqnarray}
 I_1^c&=&  (|A_{L0}|^2+|A_{R0}|^2)
 +8  \hat m_l^2 {\rm Re}[A_{L0}A^*_{R0} ]+4 \hat m_l^2  |A_t|^2, \nonumber\\
 I_1^s&=& \left(3/4- \hat m_l^2 \right)[|A_{L\perp}|^2
 +|A_{L||}|^2+|A_{R\perp}|^2+|A_{R||}|^2  ]
+ 4 \hat m_l^2  {\rm Re}[A_{L\perp}A_{R\perp}^*
 + A_{L||}A_{R||}^*],\nonumber\\
 I_2^c  &=& -\beta_l^2(  |A_{L0}|^2+ |A_{R0}|^2),\nonumber\\
 I_2^s  &=&
 \frac{1}{4}\beta_l^2(|A_{L\perp}|^2+|A_{L||}|^2+|A_{R\perp}|^2+|A_{R||}|^2),
 \nonumber\\
 I_3  &=&\frac{1}{2}\beta_l^2(|A_{L\perp}|^2-|A_{L||}|^2+|A_{R\perp}|^2
 -|A_{R||}|^2),\nonumber\\
 I_4
  &=& \frac{1}{\sqrt2}\beta_l^2
  [{\rm Re}(A_{L0}A_{L||}^*)+{\rm
  Re}(A_{R0}A_{R||}^*)],\;\;\;\;\;\;\;\;
 I_5
  = \sqrt 2\beta_l
  [{\rm Re}(A_{L0}A_{L\perp}^*)-{\rm Re}(A_{R0}A_{R\perp}^*)],\nonumber\\
 I_6  &=& 2\beta_l
  [{\rm Re}(A_{L||}A^*_{L\perp})-{\rm
  Re}(A_{R||}A^*_{R\perp})],\;\;\;\;\;
 I_7
 = \sqrt2\beta_l
  [{\rm Im}(A_{L0}A^*_{L||})-{\rm Im}(A_{R0}A^*_{R||})],\nonumber\\
 I_8 &=& \frac{1}{\sqrt2}\beta_l^2
  [{\rm Im}(A_{L0}A^*_{L\perp})+{\rm
  Im}(A_{R0}A^*_{R\perp})],\;\;\;\;\;
 I_9
 =\beta_l^2
  [{\rm Im}(A_{L||}A^*_{L\perp})+{\rm
  Im}(A_{R||}A^*_{R\perp})],\label{eq:angularCoefficients}
\end{eqnarray}
with $\hat{m}_l = m_l/\sqrt{q^2}$.  
Substituting the  expressions $A_i$ into the angular coefficients, we obtain
\begin{eqnarray}
 I_1^c&=& 
 \sum_{J=0,...}  \bigg\{ |Y_J^0|^2 \left[ |A^J_{L0}|^2+|A^J_{R0}|^2
 +8  \hat m_l^2  | A^J_{L0}A^{J*}_{R0} | \cos(\delta_{L0}^J -\delta_{R0}^J) 
 +4 \hat m_l^2  |A_t^J|^2\right]   \nonumber\\
 &&+ \sum_{ J'=J+1, ...} Y_J^0Y_{J'}^0\left[ 2\cos(\delta_{L0}^J - 
 \delta^{J'}_{L0})|A^J_{L0}||A^{J'*}_{L0}| +2\cos(\delta_{R0}^J - 
 \delta^{J'}_{R0})|A^J_{R0}||A^{J'*}_{R0}| \right. \nonumber\\
 && \left. 
 +8 \hat m_l^2 [ \cos(\delta_{L0}^J -\delta_{R0}^{J'})|A^J_{L0}A^{J'*}_{R0}| 
 + \cos(\delta_{L0}^{J'} -\delta_{R0}^{J})|A^{J'}_{L0}A^{J*}_{R0}| ]
 +8\hat m_l^2  
 \cos (\delta_{t}^J -\delta_{t}^{J'} )|A^J_t||A^{J'}_t|\right]\bigg\}, 
 \\
 I_1^s&=& \sum_{J=1,...}  \bigg\{ |Y_J^{-1}|^2 \left[    \left(3/4-   
 \hat m_l^2\right)  [|A^J_{L\perp}|^2+|A^J_{L||}|^2+|A^J_{R\perp}|^2
 +|A^J_{R||}|^2  ]\right.
   \nonumber\\ 
 && \left.
 + 4\hat m_l^2   \left( \cos(\delta_{L\perp}^J 
 -\delta_{R\perp}^{J})|A^J_{L\perp}A^J_{R\perp}|
 + \cos(\delta_{L||}^J -\delta_{R||}^{J})|A^J_{L||}A^J_{R||}|
 \right) \right]
 \nonumber\\
 && \left.+  \sum_{ J'=J+1, ...} Y_J^{-1}Y_{J'}^{-1} \left[ \left(3/4
 -   \hat m_l^2\right) [ 2\cos(\delta_{L\perp}^{J} 
 - \delta_{L\perp}^{J'})|A_{L\perp}^J||A_{L\perp}^{J'}| + (L \to R) 
 + (\perp \to ||)]  \right.\right.  \nonumber\\
 && \left.+ 4\hat m_l^2  [ \cos(\delta_{L\perp}^J 
 -\delta_{R\perp}^{J'})|A^J_{L\perp}A^{J'*}_{R\perp}| 
 + \cos(\delta_{L\perp}^{J'} 
 -\delta_{R\perp}^{J})|A^{J'}_{L\perp}A^{J*}_{R\perp}| 
 + (\perp \to ||)]\right]\bigg\},\non
 I_2^c  &=& -\beta_l^2  \sum_{J=0,...}  \bigg\{ |Y_J^0|^2  
  (|A^J_{L0}|^2+|A^J_{R0}|^2)  +\sum_{ J'=J+1, ...} Y_J^0Y_{J'}^0
  \left[ 2\cos(\delta_{L0}^J - \delta^{J'}_{L0})|A^J_{L0}||A^{J'}_{L0}| 
  +  (L\to R) \right]\bigg\} ,\non
 I_2^s  &=&
 \frac{1}{4}\beta_l^2 \sum_{J=0,...}  \bigg\{  |Y_J^{-1}|^2 
 \left[ (|A^J_{L\perp}|^2+|A^J_{L||}|^2)\right]    \nonumber\\
 &&  +\sum_{ J'=J+1} Y_J^{-1}Y_{J'}^{-1}\left[ 2\cos(\delta_{L\perp}^J 
 - \delta^{J'}_{L\perp})|A^J_{L\perp}||A^{J'}_{L\perp}| +2\cos(\delta_{L||}^J 
 - \delta^{J'}_{L||})|A^J_{L||}||A^{J'}_{L||}| \right]
 + (L\to R)\bigg\},
 \non
 I_3  &=&
 \frac{1}{2}\beta_l^2  \sum_{J=1,...}  \bigg\{ |Y_J^{-1}|^2 \left[ (|A^J_{L\perp}|^2-|A^J_{L||}|^2)\right]    \nonumber\\
 &&  +\sum_{ J'=J+1} Y_J^{-1}Y_{J'}^{-1}\left[ 2\cos(\delta_{L\perp}^J - \delta^{J'}_{L\perp})|A^J_{L\perp}||A^{J'}_{L\perp}| -2\cos(\delta_{L||}^J - \delta^{J'}_{L||})|A^J_{L||}||A^{J'}_{L||}| \right]+ (L\to R)\bigg\},
 \non
 I_4
  &=& \frac{1}{\sqrt2}\beta_l^2   \sum_{J=1, ...} \sum_{J'=1, ..} \left[  Y_J^0 Y_{J'}^{-1} | A^J_{L0}A^{J'*}_{L||} | \cos(\delta_{L0}^J -\delta_{L||}^{J'}) +(L\to R)\right],
  \non
 I_5
  &=& \sqrt 2\beta_l  \sum_{J=0, ...} \sum_{J'=1, ..} \left[   Y_J^0 Y_{J'}^{-1} | A^J_{L0}A^{J'*}_{L\perp} | \cos(\delta_{L0}^J -\delta_{L\perp}^{J'}) -(L\to R)\right],
  \non
 I_6  &=& 2\beta_l \sum_{J=1,...}  \bigg\{  |Y_J^{-1}|^2 | A^J_{L||}A^{J*}_{L\perp} | \cos(\delta_{L||}^J -\delta_{L\perp}^J) \nonumber\\
 && + \sum_{J'=J+1} Y_J^{-1}Y_{J'}^{-1} [ \cos(\delta_{L||}^J -\delta_{L\perp}^{J'})|A^J_{L||}A^{J'}_{L\perp}| + \cos(\delta_{L||}^{J'} -\delta_{L\perp}^{J})|A^{J'}_{L||}A^{J}_{L\perp}| ] -(L\to R)\bigg\},
 \non
  I_7
 &=& \sqrt2\beta_l  \sum_{J=0, ...} \sum_{J'=1, ..} \left[  Y_J^0 Y_{J'}^{-1} | A^J_{L0}A^{J'*}_{L||} | \sin(\delta_{L0}^J -\delta_{L||}^{J'}) -(L\to R)\right],
 \non
 I_8 &=& \frac{1}{\sqrt2}\beta_l^2 \sum_{J=0, ...} \sum_{J'=1, ..} \left[  Y_J^0 Y_{J'}^{-1} | A^J_{L0}A^{J'*}_{L\perp} | \sin(\delta_{L0}^J -\delta_{L\perp}^{J'}) +(L\to R)\right],
 \non
 I_9
 &=&\beta_l^2\sum_{J=1,...}  \bigg\{ |Y_J^{-1}|^2 | A^J_{L||}A^{J}_{L\perp} | \sin(\delta_{L||}^J -\delta_{L\perp}^J) \nonumber\\
 && + \sum_{J'=J+1} Y_J^{-1}Y_{J'}^{-1} [ \sin(\delta_{L||}^J -\delta_{L\perp}^{J'})|A^J_{L||}A^{J'}_{L\perp}| + \sin(\delta_{L||}^{J'} -\delta_{L\perp}^{J})|A^{J'}_{L||}A^{J}_{L\perp}| ] +(L\to R)  \bigg\},
 \label{eq:simplified_angularCoefficients}
\end{eqnarray}
where for brevity we have omitted the argument in the spherical harmonics
functions: $Y_J^i\equiv  Y_J^i(\theta_K, 0)$.  Transverse amplitudes vanish for
$J=0$ since spin-0 mesons have only one polarization configuration. 


\section{Normalization of the Line-Shape}
\label{sec:normalization}

Assuming the matrix elements  to be saturated by  resonances, we
have 
\begin{eqnarray}
 \langle K\pi | [\bar s b] |\overline B^0\rangle &\simeq &  \int d^4 p_{K^*_J} \frac{ \langle K\pi|K^*_J\rangle \langle K^*_J | [\bar s b] |\overline B^0\rangle }{ p_{K^*_J}^2- m_{K^*_J}^2 +i m_{K^*_J} \Gamma_{K^*_J}} \sim \frac{ \langle K^*_J | [\bar s b] |\overline B^0\rangle    \langle K\pi |\bar s u|0\rangle }{\langle K^*_J |\bar su |0\rangle  } 
 \end{eqnarray}
 and 
 \begin{eqnarray}
  \langle K\pi |\bar s u|0\rangle \simeq \int d^4 p_{K^*_J} \frac{ \langle K\pi |K^*_J\rangle \langle K^*_J |\bar su |0\rangle   }{ p_{K^*_J}^2 -m_{K^*_J}^2+i m_{K^*_J} \Gamma_{K^*_J}}.
 \end{eqnarray}
By comparing these two equations, we derive the line-shape used in this work:
\begin{eqnarray}
 L_{K^*_0}(m_{K\pi})= \sqrt { \frac{|\vec p\,|}{8\pi m_{K_{0}^*}}} \frac{\langle K\pi |\bar s u|0\rangle}{\langle K_{0}^* |\bar s u|0\rangle}.
\end{eqnarray}
Although this equation  can not be taken literally, we determine the
normalization constant ${\cal N}_{\rm \chi PT}$ from this equation,  bearing in
mind the large uncertainties.  It is important to notice that below threshold
the momentum becomes negative and thus an extra imaginary part is introduced.
This is not appropriate in the case of $K\bar K$, and thus we   use the Flatt\'e
model instead. 
 
\end{appendix}




\begin{thebibliography}{11}


\bibitem{Lees:2012tva} 
  J.~P.~Lees {\it et al.}  [BaBar Collaboration],
  Phys.\ Rev.\ D {\bf 86}, 032012 (2012)
  [arXiv:1204.3933 [hep-ex]].


\bibitem{Wei:2009zv} 
  J.-T.~Wei {\it et al.}  [BELLE Collaboration],
  Phys.\ Rev.\ Lett.\  {\bf 103}, 171801 (2009)
  [arXiv:0904.0770 [hep-ex]].


\bibitem{Aaltonen:2011cn} 
  T.~Aaltonen {\it et al.}  [CDF Collaboration],
  Phys.\ Rev.\ Lett.\  {\bf 106}, 161801 (2011)
  [arXiv:1101.1028 [hep-ex]].


\bibitem{Aaij:2013iag} 
  R.~Aaij {\it et al.}  [LHCb Collaboration],
  arXiv:1304.6325 [hep-ex].


\bibitem{Ali:1999mm} 
  A.~Ali, P.~Ball, L.~T.~Handoko and G.~Hiller,
  Phys.\ Rev.\ D {\bf 61}, 074024 (2000)
  [hep-ph/9910221].


\bibitem{Chen:2002bq} 
  C.-H.~Chen and C.~Q.~Geng,
  Nucl.\ Phys.\ B {\bf 636}, 338 (2002)
  [hep-ph/0203003].


\bibitem{Kruger:2005ep} 
  F.~Kr\"uger and J.~Matias,
  Phys.\ Rev.\ D {\bf 71}, 094009 (2005)
  [hep-ph/0502060].


\bibitem{Egede:2008uy} 
  U.~Egede, T.~Hurth, J.~Matias, M.~Ramon and W.~Reece,
  JHEP {\bf 0811}, 032 (2008)
  [arXiv:0807.2589 [hep-ph]].


\bibitem{Altmannshofer:2008dz} 
  W.~Altmannshofer, P.~Ball, A.~Bharucha, A.~J.~Buras, D.~M.~Straub and M.~Wick,
  JHEP {\bf 0901}, 019 (2009)
  [arXiv:0811.1214 [hep-ph]].


\bibitem{Chiang:2009dx} 
  C.-W.~Chiang, R.-H.~Li and C.-D.~Lu,
  Chin.\ Phys.\ C {\bf 36}, 14 (2012)
  [arXiv:0911.2399 [hep-ph]].


\bibitem{Khodjamirian:2010vf} 
  A.~Khodjamirian, T.~.Mannel, A.~A.~Pivovarov and Y.~-M.~Wang,
  JHEP {\bf 1009}, 089 (2010)
  [arXiv:1006.4945 [hep-ph]].


\bibitem{Bobeth:2011gi} 
  C.~Bobeth, G.~Hiller and D.~van Dyk,
  JHEP {\bf 1107}, 067 (2011)
  [arXiv:1105.0376 [hep-ph]].


\bibitem{Bobeth:2012vn} 
  C.~Bobeth, G.~Hiller and D.~van Dyk,
  Phys.\ Rev.\ D {\bf 87}, 034016 (2013)
  [arXiv:1212.2321 [hep-ph]].


\bibitem{Jager:2012uw} 
  S.~J\"ager and J.~M.~Camalich,
  JHEP {\bf 1305}, 043 (2013)
  [arXiv:1212.2263 [hep-ph]].


\bibitem{Descotes-Genon:2013vna} 
  S.~Descotes-Genon, T.~Hurth, J.~Matias and J.~Virto,
  JHEP {\bf 1305}, 137 (2013)
  [arXiv:1303.5794 [hep-ph]].


\bibitem{Lu:2011jm} 
  C.-D.~Lu and W.~Wang,
  Phys.\ Rev.\ D {\bf 85}, 034014 (2012)
  [arXiv:1111.1513 [hep-ph]].


\bibitem{Li:2010ra} 
  R.-H.~Li, C.-D.~Lu and W.~Wang,
  Phys.\ Rev.\ D {\bf 83}, 034034 (2011)
  [arXiv:1012.2129 [hep-ph]].


\bibitem{Becirevic:2012dp} 
  D.~Becirevic and A.~Tayduganov,
  Nucl.\ Phys.\ B {\bf 868}, 368 (2013)
  [arXiv:1207.4004 [hep-ph]].


\bibitem{Matias:2012qz} 
  J.~Matias,
  Phys.\ Rev.\ D {\bf 86}, 094024 (2012)
  [arXiv:1209.1525 [hep-ph]].


\bibitem{Blake:2012mb} 
  T.~Blake, U.~Egede and A.~Shires,
  JHEP {\bf 1303}, 027 (2013)
  [arXiv:1210.5279 [hep-ph]].


\bibitem{delAmoSanchez:2010fd} 
  P.~del Amo Sanchez {\it et al.}  [BaBar Collaboration],
  Phys.\ Rev.\ D {\bf 83}, 072001 (2011)
  [arXiv:1012.1810 [hep-ex]].

\bibitem{Link:2009ng} 
  J.~M.~Link {\it et al.}  [FOCUS Collaboration],
  Phys.\ Lett.\ B {\bf 681}, 14 (2009)
  [arXiv:0905.4846 [hep-ex]].

\bibitem{Aitala:2005yh} 
  E.~M.~Aitala {\it et al.}  [E791 Collaboration],
  Phys.\ Rev.\ D {\bf 73}, 032004 (2006)
  [Erratum-ibid.\ D {\bf 74}, 059901 (2006)]
  [hep-ex/0507099].

\bibitem{Oller:2004xm} 
  J.~A.~Oller,
  Phys.\ Rev.\ D {\bf 71}, 054030 (2005)
  [hep-ph/0411105].

\bibitem{Bediaga:2004bc} 
  I.~Bediaga and J.~M.~de Miranda,
  Phys.\ Lett.\ B {\bf 633}, 167 (2006)
  [hep-ex/0405019].


\bibitem{Magalhaes:2011sh} 
  P.~C.~Magalhaes, M.~R.~Robilotta, K.~S.~F.~F.~Guimaraes, T.~Frederico, W.~de Paula, I.~Bediaga, A.~C.~d.~Reis and C.~M.~Maekawa {\it et al.},
  Phys.\ Rev.\ D {\bf 84}, 094001 (2011)
  [arXiv:1105.5120 [hep-ph]].


\bibitem{Kamano:2011ih} 
  H.~Kamano, S.~X.~Nakamura, T.~S.~H.~Lee and T.~Sato,
  Phys.\ Rev.\ D {\bf 84}, 114019 (2011)
  [arXiv:1106.4523 [hep-ph]].

\bibitem{Bugg:2005ni} 
  D.~V.~Bugg,
  Eur.\ Phys.\ J.\ A {\bf 25}, 107 (2005)
  [Erratum-ibid.\ A {\bf 26}, 151 (2005)]
  [hep-ex/0510026].


\bibitem{Aston:1987ir} 
  D.~Aston, N.~Awaji, T.~Bienz, F.~Bird, J.~D'Amore, W.~M.~Dunwoodie, R.~Endorf and K.~Fujii {\it et al.},
  Nucl.\ Phys.\ B {\bf 296}, 493 (1988).


\bibitem{Lang:2012sv} 
  C.~B.~Lang, L.~Leskovec, D.~Mohler and S.~Prelovsek,
  Phys.\ Rev.\ D {\bf 86}, 054508 (2012)
  [arXiv:1207.3204 [hep-lat]].

\bibitem{Beane:2006gj} 
  S.~R.~Beane, P.~F.~Bedaque, T.~C.~Luu, K.~Orginos, E.~Pallante, A.~Parreno and M.~J.~Savage,
  Phys.\ Rev.\ D {\bf 74}, 114503 (2006)
  [hep-lat/0607036].

\bibitem{Doring:2011nd} 
  M.~D\"oring and U.-G.~Mei{\ss}ner,
  JHEP {\bf 1201}, 009 (2012)
  [arXiv:1111.0616 [hep-lat]].

\bibitem{Doring:2012eu} 
  M.~D\"oring, U.-G.~Mei{\ss}ner, E.~Oset and A.~Rusetsky,
  Eur.\ Phys.\ J.\ A {\bf 48}, 114 (2012)
  [arXiv:1205.4838 [hep-lat]].

\bibitem{Doring:2011vk} 
  M.~D\"oring, U.-G.~Mei{\ss}ner, E.~Oset and A.~Rusetsky,
  Eur.\ Phys.\ J.\ A {\bf 47}, 139 (2011)
  [arXiv:1107.3988 [hep-lat]].


\bibitem{Aubert:2008rs} 
  B.~Aubert {\it et al.}  [BaBar Collaboration],
  Phys.\ Rev.\ D {\bf 78}, 051101 (2008)
  [arXiv:0807.1599 [hep-ex]].


\bibitem{Ecklund:2009aa} 
  K.~M.~Ecklund {\it et al.}  [CLEO Collaboration],
  Phys.\ Rev.\ D {\bf 80}, 052009 (2009)
  [arXiv:0907.3201 [hep-ex]].


\bibitem{Gardner:2001gc} 
  S.~Gardner and U.-G.~Mei{\ss}ner,
  Phys.\ Rev.\ D {\bf 65}, 094004 (2002)
  [hep-ph/0112281].



\bibitem{Keum:2000ph} 
  Y.-Y.~Keum, H.-N.~Li and A.~I.~Sanda,
  Phys.\ Lett.\ B {\bf 504}, 6 (2001)
  [hep-ph/0004004].


\bibitem{Keum:2000wi} 
  Y.~Y.~Keum, H.-N.~Li and A.~I.~Sanda,
  Phys.\ Rev.\ D {\bf 63}, 054008 (2001)
  [hep-ph/0004173].


\bibitem{Lu:2000em} 
  C.-D.~Lu, K.~Ukai and M.-Z.~Yang,
  Phys.\ Rev.\ D {\bf 63}, 074009 (2001)
  [hep-ph/0004213].


\bibitem{Lu:2000hj} 
  C.-D.~Lu and M.-Z.~Yang,
  Eur.\ Phys.\ J.\ C {\bf 23}, 275 (2002)
  [hep-ph/0011238].


\bibitem{Li:2012nk} 
  H.-N.~Li, Y.-L.~Shen and Y.-M.~Wang,
  Phys.\ Rev.\ D {\bf 85}, 074004 (2012)
  [arXiv:1201.5066 [hep-ph]].


\bibitem{Beringer:1900zz} 
  J.~Beringer {\it et al.}  [Particle Data Group Collaboration],
  Phys.\ Rev.\ D {\bf 86}, 010001 (2012).


\bibitem{Gasser:1990bv} 
  J.~Gasser and U.-G.~Mei{\ss}ner,
  Nucl.\ Phys.\ B {\bf 357}, 90 (1991).


\bibitem{Meissner:2000bc} 
  U.-G.~Mei{\ss}ner and J.~A.~Oller,
  Nucl.\ Phys.\ A {\bf 679}, 671 (2001)
  [hep-ph/0005253].


\bibitem{Oller:2000ug} 
  J.~A.~Oller, E.~Oset and J.~E.~Palomar,
  Phys.\ Rev.\ D {\bf 63}, 114009 (2001)
  [hep-ph/0011096].


\bibitem{Frink:2002ht} 
  M.~Frink, B.~Kubis and U.-G.~Mei{\ss}ner,
  Eur.\ Phys.\ J.\ C {\bf 25}, 259 (2002)
  [hep-ph/0203193].


\bibitem{Bijnens:2003uy} 
  J.~Bijnens and P.~Talavera,
  Nucl.\ Phys.\ B {\bf 669}, 341 (2003)
  [hep-ph/0303103].


\bibitem{Lahde:2006wr} 
  T.~A.~L\"ahde and U.-G.~Mei{\ss}ner,
  Phys.\ Rev.\ D {\bf 74}, 034021 (2006)
  [hep-ph/0606133].


\bibitem{Bernard:2009ds} 
  V.~Bernard and E.~Passemar,
  JHEP {\bf 1004}, 001 (2010)
  [arXiv:0912.3792 [hep-ph]].


\bibitem{Guo:2012yt} 
  Z.-H.~Guo, J.~A.~Oller and J.~Ruiz de Elvira,
  Phys.\ Rev.\ D {\bf 86}, 054006 (2012)
  [arXiv:1206.4163 [hep-ph]].


\bibitem{Donoghue:1990xh} 
  J.~F.~Donoghue, J.~Gasser and H.~Leutwyler,
  Nucl.\ Phys.\ B {\bf 343}, 341 (1990).


\bibitem{Jamin:2000wn} 
  M.~Jamin, J.~A.~Oller and A.~Pich,
  Nucl.\ Phys.\ B {\bf 587}, 331 (2000)
  [hep-ph/0006045].


\bibitem{Jamin:2001zq} 
  M.~Jamin, J.~A.~Oller and A.~Pich,
  Nucl.\ Phys.\ B {\bf 622}, 279 (2002)
  [hep-ph/0110193].


\bibitem{Jamin:2006tj} 
  M.~Jamin, J.~A.~Oller and A.~Pich,
  Phys.\ Rev.\ D {\bf 74}, 074009 (2006)
  [hep-ph/0605095].


\bibitem{Bernard:2007tk} 
  V.~Bernard and E.~Passemar,
  Phys.\ Lett.\ B {\bf 661}, 95 (2008)
  [arXiv:0711.3450 [hep-ph]].


\bibitem{Guerrero:1997ku} 
  F.~Guerrero and A.~Pich,
  Phys.\ Lett.\ B {\bf 412}, 382 (1997)
  [hep-ph/9707347].


\bibitem{Guo:2008nc} 
  F.-K.~Guo, C.~Hanhart, F.~J.~Llanes-Estrada and U.-G.~Mei{\ss}ner,
  Phys.\ Lett.\ B {\bf 678}, 90 (2009)
  [arXiv:0812.3270 [hep-ph]].


\bibitem{Hanhart:2012wi} 
  C.~Hanhart,
  Phys.\ Lett.\ B {\bf 715}, 170 (2012)
  [arXiv:1203.6839 [hep-ph]].


\bibitem{Bernard:2011ae} 
  V.~Bernard, D.~R.~Boito and E.~Passemar,
  Nucl.\ Phys.\ Proc.\ Suppl.\  {\bf 218}, 140 (2011)
  [arXiv:1103.4855 [hep-ph]].


\bibitem{Oller:1998hw} 
  J.~A.~Oller, E.~Oset and J.~R.~Pelaez,
  Phys.\ Rev.\ D {\bf 59}, 074001 (1999)
  [Erratum-ibid.\ D {\bf 60}, 099906 (1999)]
  [Erratum-ibid.\ D {\bf 75}, 099903 (2007)]
  [hep-ph/9804209].


\bibitem{Guo:2011pa} 
  Z.-H.~Guo and J.~A.~Oller,
  Phys.\ Rev.\ D {\bf 84}, 034005 (2011)
  [arXiv:1104.2849 [hep-ph]].


\bibitem{Albaladejo:2008qa} 
  M.~Albaladejo and J.~A.~Oller,
  Phys.\ Rev.\ Lett.\  {\bf 101}, 252002 (2008)
  [arXiv:0801.4929 [hep-ph]].


\bibitem{GomezNicola:2001as} 
  A.~Gomez Nicola and J.~R.~Pelaez,
  Phys.\ Rev.\ D {\bf 65}, 054009 (2002)
  [hep-ph/0109056].

\bibitem{Bernard:1991zc}
  V.~Bernard, N.~Kaiser and U.-G.~Mei{\ss}ner,
  Nucl.\ Phys.\ B {\bf 364} (1991) 283.


\bibitem{DescotesGenon:2006uk} 
  S.~Descotes-Genon and B.~Moussallam,
  Eur.\ Phys.\ J.\ C {\bf 48}, 553 (2006)
  [hep-ph/0607133].


\bibitem{Bugg:2009uk} 
  D.~V.~Bugg,
  Phys.\ Rev.\ D {\bf 81}, 014002 (2010)
  [arXiv:0906.3992 [hep-ph]].


\bibitem{Buettiker:2003pp} 
  P.~B\"uttiker, S.~Descotes-Genon and B.~Moussallam,
  Eur.\ Phys.\ J.\ C {\bf 33}, 409 (2004)
  [hep-ph/0310283].


\bibitem{DescotesGenon:2007ta} 
  S.~Descotes-Genon,
  Eur.\ Phys.\ J.\ C {\bf 52}, 141 (2007)
  [hep-ph/0703154 [HEP-PH]].


\bibitem{Gasser:1984ux} 
  J.~Gasser and H.~Leutwyler,
  Nucl.\ Phys.\ B {\bf 250}, 517 (1985).


\bibitem{Bijnens:2011tb} 
  J.~Bijnens and I.~Jemos,
  Nucl.\ Phys.\ B {\bf 854}, 631 (2012)
  [arXiv:1103.5945 [hep-ph]].


\bibitem{Gasser:1984gg} 
  J.~Gasser and H.~Leutwyler,
  Nucl.\ Phys.\ B {\bf 250}, 465 (1985).


\bibitem{Niecknig:2012sj} 
  F.~Niecknig, B.~Kubis and S.~P.~Schneider,
  Eur.\ Phys.\ J.\ C {\bf 72}, 2014 (2012)
  [arXiv:1203.2501 [hep-ph]].

\bibitem{Haftel:1970zz}
  M.~I.~Haftel and F.~Tabakin,
  Nucl.\ Phys.\ A {\bf 158}, 1 (1970).


\bibitem{Hoferichter:2012wf} 
  M.~Hoferichter, C.~Ditsche, B.~Kubis and U.-G.~Mei{\ss}ner,
  JHEP {\bf 1206}, 063 (2012)
  [arXiv:1204.6251 [hep-ph]].


\bibitem{Oller:2007xd} 
  J.~A.~Oller and L.~Roca,
  Phys.\ Lett.\ B {\bf 651}, 139 (2007)
  [arXiv:0704.0039 [hep-ph]].


\bibitem{Gasser:1983yg} 
  J.~Gasser and H.~Leutwyler,
  Annals Phys.\  {\bf 158}, 142 (1984).


\bibitem{Oller:1997ti} 
  J.~A.~Oller and E.~Oset,
  Nucl.\ Phys.\ A {\bf 620}, 438 (1997)
  [Erratum-ibid.\ A {\bf 652}, 407 (1999)]
  [hep-ph/9702314].


\bibitem{Aubert:2005ce} 
  B.~Aubert {\it et al.}  [BaBar Collaboration],
  Phys.\ Rev.\ D {\bf 72}, 072003 (2005)
  [Erratum-ibid.\ D {\bf 74}, 099903 (2006)]
  [hep-ex/0507004].


\bibitem{Aubert:2008aa} 
  B.~Aubert {\it et al.}  [BaBar Collaboration],
  Phys.\ Rev.\ D {\bf 79}, 112001 (2009)
  [arXiv:0811.0564 [hep-ex]].


\bibitem{Flatte:1976xu} 
  S.~M.~Flatt\'e,
  Phys.\ Lett.\ B {\bf 63}, 224 (1976).


\bibitem{Ablikim:2004wn} 
  M.~Ablikim {\it et al.}  [BES Collaboration],
  Phys.\ Lett.\ B {\bf 607}, 243 (2005)
  [hep-ex/0411001].


\bibitem{Aaij:2013aln} 
  RAaij {\it et al.}  [ LHCb Collaboration],
  arXiv:1305.2168 [hep-ex].


\bibitem{Charles:1998dr} 
  J.~Charles, A.~Le Yaouanc, L.~Oliver, O.~Pene and J.~C.~Raynal,
  Phys.\ Rev.\ D {\bf 60}, 014001 (1999)
  [hep-ph/9812358].


\bibitem{Diehl:2005rn} 
  M.~Diehl, A.~Manashov and A.~Sch\"afer,
  Phys.\ Lett.\ B {\bf 622}, 69 (2005)
  [hep-ph/0505269].

\bibitem{Chen:2002th} 
  C.-H.~Chen and H.-N.~Li,
  Phys.\ Lett.\ B {\bf 561}, 258 (2003)
  [hep-ph/0209043].

\bibitem{Cheng:2005ug} 
  H.-Y.~Cheng, C.-K.~Chua and A.~Soni,
  Phys.\ Rev.\ D {\bf 72}, 094003 (2005)
  [hep-ph/0506268].
  
\bibitem{ElBennich:2009da} 
  B.~El-Bennich, A.~Furman, R.~Kaminski, L.~Lesniak, B.~Loiseau and B.~Moussallam,
  Phys.\ Rev.\ D {\bf 79}, 094005 (2009)
  [Erratum-ibid.\ D {\bf 83}, 039903 (2011)]
  [arXiv:0902.3645 [hep-ph]].

\bibitem{Hatanaka:2009sj} 
  H.~Hatanaka and K.-C.~Yang,
  Eur.\ Phys.\ J.\ C {\bf 67}, 149 (2010)
  [arXiv:0907.1496 [hep-ph]].


\bibitem{Wang:2010ni} 
  W.~Wang,
  Phys.\ Rev.\ D {\bf 83}, 014008 (2011)
  [arXiv:1008.5326 [hep-ph]].


\bibitem{Yang:2010qd} 
  K.-C.~Yang,
  Phys.\ Lett.\ B {\bf 695}, 444 (2011)
  [arXiv:1010.2944 [hep-ph]].


\bibitem{Li:2008tk} 
  R.-H.~Li, C.-D.~Lu, W.~Wang and X.-X.~Wang,
  Phys.\ Rev.\ D {\bf 79}, 014013 (2009)
  [arXiv:0811.2648 [hep-ph]].


\bibitem{Wang:2006ria} 
  W.~Wang, Y.-L.~Shen, Y.~Li and C.-D.~Lu,
  Phys.\ Rev.\ D {\bf 74}, 114010 (2006)
  [hep-ph/0609082].


\bibitem{Shen:2006ms} 
  Y.-L.~Shen, W.~Wang, J.~Zhu and C.-D.~Lu,
  Eur.\ Phys.\ J.\ C {\bf 50}, 877 (2007)
  [hep-ph/0610380].


\bibitem{Kim:2009dg} 
  C.~S.~Kim, Y.~Li and W.~Wang,
  Phys.\ Rev.\ D {\bf 81}, 074014 (2010)
  [arXiv:0912.1718 [hep-ph]].

\bibitem{Cheng:2005nb} 
  H.-Y.~Cheng, C.-K.~Chua and K.-C.~Yang,
  Phys.\ Rev.\ D {\bf 73}, 014017 (2006)
  [hep-ph/0508104].

\bibitem{Ball:2004rg} 
  P.~Ball and R.~Zwicky,
  Phys.\ Rev.\ D {\bf 71}, 014029 (2005)
  [hep-ph/0412079].


\bibitem{Khodjamirian:2006st} 
  A.~Khodjamirian, T.~Mannel and N.~Offen,
  Phys.\ Rev.\ D {\bf 75}, 054013 (2007)
  [hep-ph/0611193].


\bibitem{Colangelo:2010bg} 
  P.~Colangelo, F.~De Fazio and W.~Wang,
  Phys.\ Rev.\ D {\bf 81}, 074001 (2010)
  [arXiv:1002.2880 [hep-ph]].

\bibitem{Colangelo:2010wg} 
  P.~Colangelo, F.~De Fazio and W.~Wang,
  Phys.\ Rev.\ D {\bf 83}, 094027 (2011)
  [arXiv:1009.4612 [hep-ph]].
  
\end{thebibliography}
\end{document}